\documentclass[12pt]{article}
\usepackage{upgreek}
\usepackage{graphicx}
\usepackage{amssymb}
\usepackage{amsmath}
\usepackage{bm}
\usepackage{indentfirst}
\usepackage{cite}

\allowdisplaybreaks

\begin{document}
\title{Dissociation cross sections of large-momentum charmonia 
with light mesons in hadronic matter}
\author{Yi-Hao Pan$^{1}$, Wen-Hao Shi$^{1}$, Xiao-Ming Xu$^{1}$ and 
H. J. Weber$^{2}$}
\date{}
\maketitle \vspace{-1cm}
\centerline{$^{1}$Department of Physics, Shanghai University, Baoshan,
Shanghai 200444, China}
\centerline{$^{2}$Department of Physics, University of Virginia, 
Charlottesville, VA 22904, USA}

\begin{abstract}
Momenta of charmonia created in Pb-Pb collisions at the CERN
Large Hadron Collider
are so large that three or more mesons may be produced when the charmonia
collide with light mesons in hadronic matter. We study the meson-charmonium
collision in a mechanism where the collision
produces two quarks and two antiquarks; the charm quark then fragments into
charmed mesons, and the other three constituents as well as quarks and
antiquarks created from vacuum give rise to two or more mesons.
The absolute square of the
transition amplitude for the production of two quarks and two 
antiquarks is derived from the $S$-matrix element, and cross-section formulas
are derived from the absolute square of the transition amplitude and 
charm-quark
fragmentation functions. With a temperature-dependent quark potential, we 
calculate unpolarized cross sections for inclusive
$D^+$, $D^0$, $D^+_s$, or $D^{*+}$ production in scattering of 
charmonia by $\pi$, $\rho$, $K$, or $K^\ast$ mesons. At low center-of-mass 
energies of the charmonium and the light meson, the cross sections are very
small. At high energies the cross sections have obvious temperature dependence,
and are comparable to peak cross sections of two-to-two meson-charmonium 
reactions.
\end{abstract}

\noindent
Keywords: Charmonium dissociation, Relativistic constituent quark potential 
model

\noindent
PACS: 13.75.Lb; 12.39.Jh; 12.39.Pn

\vspace{0.5cm}
\leftline{\bf 1. Introduction}
\vspace{0.5cm}

Plenty of efforts with quantum chromodynamics (QCD) and effective field 
theories have been devoted to exploring strong interactions of charmonia with 
light 
hadrons. Four main approaches involved in the study are the short-distance
approach, the effective meson approach, the quark-interchange approach, and
QCD sum rules. In the short-distance approach, methods in perturbative QCD,
for example, the operator product expansion, were applied to charmonia
of small sizes
\cite{KS,Peskin,AGGA}. In the effective meson approach, effective meson 
Lagrangians with different symmetries and Feynman diagrams with various vertex
functions have been used to get millibarn-scale cross sections for 
meson-charmonium reactions
\cite{MM,Haglin,LK,OSL,NNR,MPPR,AN,BG,CMR,AKM,ACM}. 
Recently, $\pi J/\psi$, $\rho \eta_c$, and $D\bar{D}^*$ ($D^*\bar{D}$) 
reactions have been put together to form coupled channels. The corresponding 
scattering amplitude 
derived in $SU(4)$ chiral perturbation theory is unitarized \cite{CMR,AKM,ACM},
and cross sections obtained for $\pi J/\psi$ dissociation
are different from those given in Refs. 
\cite{MM,Haglin,LK,OSL,NNR,MPPR,AN,BG}.
In the quark-interchange approach, charmonium dissociation in collisions with
light mesons is caused by quark interchange \cite{BS} between the charmonium
and the meson. The dissociation has been studied in the Born approximation 
with quark potentials which reproduce spectroscopic data and provide mesonic
quark-antiquark wave functions
\cite{MBQ,WSB,BSWX,ZX,JSX,LJX}. In the QCD sum rules, $\pi J/\psi$
dissociation cross sections were obtained from the general vacuum-pion
correlation function with the currents of $J/\psi$ and of charmed mesons in the
soft-pion limit \cite{DKLNN}.

The studies reported in Refs. \cite{KS,Peskin,AGGA,MM,Haglin,LK,OSL,NNR,MPPR,
AN,BG,CMR,AKM,ACM,MBQ,WSB,BSWX,DKLNN} concentrate on charmonium 
dissociation
in vacuum. In hadronic matter the dissociation is different \cite{Wong}.
From perturbative QCD and lattice QCD, a temperature-dependent quark potential
was derived in Refs. \cite{ZX,Xu2002}. This potential leads to temperature 
dependence of meson masses, 
mesonic quark-antiquark relative-motion wave functions, and dissociation
cross sections \cite{ZX,JSX,LJX}. Adopting temperature dependence in meson
masses and two-meson Green functions in the coupled-channel unitary approach, 
temperature-dependent cross sections
for $J/\psi$ scattering by light mesons have been obtained in Ref. \cite{AV}.

For Pb-Pb collisions at the center-of-mass energy per nucleon-nucleon pair
$\sqrt{s_{NN}}=$ 5.02 TeV at the CERN
Large Hadron Collider, the prompt-$J/\psi$ 
transverse momentum measured by the CMS Collaboration \cite{CMS} and the ATLAS 
Collaboration \cite{ATLAS} goes up to 
50 GeV/$c$. $J/\psi$ mesons with such large transverse momenta may be broken up
due to collisions with light mesons in hadronic matter, and three or more 
mesons can be produced. Since the reactions studied
in Refs. \cite{MM,Haglin,LK,OSL,NNR,MPPR,AN,BG,CMR,AKM,ACM,MBQ,WSB,BSWX,
ZX,JSX,LJX,DKLNN,AV} were limited to two-to-two meson-charmonium reactions,  
a mechanism for large-momentum charmonium dissociation is proposed in Ref. 
\cite{JXH} to study the production of three or more mesons
in pion-charmonium collisions. 
In this mechanism a collision between a light meson and a charmonium produces 
quarks and antiquarks first, the charm quark then fragments into charmed
hadrons, and finally three or more mesons are produced. 
Besides pions, $\rho$ mesons, kaons, vector kaons, and so on in hadronic 
matter also induce charmonium dissociation. This motivates studying the 
production of three or more mesons in the dissociation of large-momentum
charmonia in collisions with light mesons in the present work.

This paper is organized as follows. In the next section we derive
cross-section formulas for dissociation of large-momentum charmonia in 
collisions with light mesons. Numerical results and relevant discussions are 
presented in Section 3. A summary is in the last section.

\vspace{0.5cm}
\leftline{\bf 2. Formalism}
\vspace{0.5cm}

We obtain the cross section for $A(q_{1}\bar{q}_{2})+B(c\bar{c})\to q_{1}
+\bar{q}_{2}+c+\bar{c}\to H_c+X$ from the cross section for 
$A+B\to q_{1}+\bar{q}_{2}+c+\bar{c}$ and the fragmentation function for 
$c\to H_c$, where $H_c$ represents a hadron that contains the charm quark. 
The fragmentation of the charm quark into hadrons is related to
quark-antiquark pairs created from the color field around the
charm quark, which is that scenario of Feynman and Field \cite{FF}. 
The charm quark may combine the antiquark of a quark-antiquark pair to form
meson $H_c$. Two quark-antiquark pairs may also form another $H_c$. All
possible ways to form meson $H_c$ are accounted for by the $c \to H_c$ 
fragmentation function. Hence, only one $H_c$ symbol is in 
$A(q_{1}\bar{q}_{2})+B(c\bar{c})\to q_{1} +\bar{q}_{2}+c+\bar{c}\to H_c+X$. 
Unused 
quarks and antiquarks combine $q_1$, $\bar{q}_2$, and $\bar c$ to form two
or more mesons. The symbol $X$ indicates the two or more mesons that do not
include meson $H_c$. We first derive cross-section formulas 
for $A+B\to q_{1}+\bar{q}_{2}+c+\bar{c}$. Let 
$E_{\rm{i}}$ be the total energy of mesons $A$ and $B$, and $E_{\rm{f}}$ be the
one of constituents $q_1$, $\bar{q}_2$, $c$, and $\bar{c}$. The $S$-matrix 
element for $A+B\to q_{1}+\bar{q}_{2}+c+\bar{c}$ is
\begin{eqnarray}
S_{\rm{fi}} &=& \delta_{\rm{fi}}-2\pi i\delta(E_{\rm{f}}-E_{\rm{i}})
(\left <q_1,\bar{q}_2,c,\bar{c}|V_{q_{1}\bar{c}}|A,B\right >
+\left <q_1,\bar{q}_2,c,\bar{c}|V_{\bar{q}_{2}c}|A,B\right >\nonumber\\
&& + \left <q_1,\bar{q}_2,c,\bar{c}|V_{q_{1}c}|A,B\right >
+\left <q_1,\bar{q}_2,c,\bar{c}|V_{\bar{q}_{2}\bar{c}}|A,B\right >),
\label{eq:sm}
\end{eqnarray}
where $V_{ab}$ is the potential between constituents $a$ and $b$. Let 
$\vec{r}_{ab}$ be the relative coordinate of $a$ and $b$, and denote the 
momentum and the position vector of meson 
$A$ ($B$) by $\vec{P}_A$ ($\vec{P}_B$) and $\vec{R}_A$ ($\vec{R}_B$), 
respectively. The wave function $|A,B\rangle$ of $A$ and $B$ is
\begin{equation}
\psi_{AB}=\frac{e^{i\vec{P}_{A}\cdot \vec{R}_{A}}}{\sqrt{V}}
\psi_{A}(\vec{r}_{q_{1}\bar{q}_{2}})
\frac{e^{i\vec{P}_{B}\cdot \vec{R}_{B}}}{\sqrt{V}}
\psi_{B}(\vec{r}_{c\bar{c}}),
\label{eq:wfAB}
\end{equation}
where every meson wave function is normalized in the volume $V$, and $\psi_A$
($\psi_B$) is a wave function of color, flavor, spin, and relative motion of 
the quark and the antiquark in meson $A$ ($B$). Let 
$\vec{p}_{q_1}^{~\prime}$ $(\vec{p}_{\bar{q}_2}^{~\prime},~\vec{p}_c^{~\prime},
~\vec{p}_{\bar{c}}^{~\prime})$ and $\vec{r}_{q_1}$ ($\vec{r}_{\bar{q}_2},
~\vec{r}_c,~\vec{r}_{\bar{c}}$) denote the momentum and the position vector of
$q_1$ $(\bar{q}_2,~c,~\bar{c})$, respectively. The wave function 
$|q_1, \bar{q}_2, c, \bar{c}\rangle$ of $q_1$, $\bar{q}_2$, $c$, and 
$\bar{c}$ is
\begin{equation}
\psi_{q_1\bar{q}_2c\bar{c}}=
\frac{e^{i\vec{p}_{q_1}^{~\prime}\cdot\vec{r}_{q_1}}}{\sqrt{V}}
\frac{e^{i\vec{p}_{\bar{q}_2}^{~\prime}\cdot\vec{r}_{\bar{q}_2}}}{\sqrt{V}}
\frac{e^{i\vec{p}_{c}^{~\prime}\cdot\vec{r}_{c}}}{\sqrt{V}}
\frac{e^{i\vec{p}_{\bar{c}}^{~\prime}\cdot\vec{r}_{\bar{c}}}}{\sqrt{V}}
\varphi_{q_1\bar{q}_2c\bar{c}{\rm color}}
\varphi_{q_1\bar{q}_2c\bar{c}{\rm flavor}}
\varphi_{q_1\bar{q}_2c\bar{c}{\rm spin}},
\label{eq:wf}
\end{equation}
where $\varphi_{q_1\bar{q}_2c\bar{c}{\rm color}}$, 
$\varphi_{q_1\bar{q}_2c\bar{c}{\rm flavor}}$, and 
$\varphi_{q_1\bar{q}_2c\bar{c}{\rm spin}}$ are the color wave function, the 
flavor wave function, and the spin wave function of $q_1$, $\bar{q}_2$, $c$, 
and $\bar{c}$, respectively. Using the wave functions, we get
\begin{eqnarray}
\left <q_1,\bar{q}_2,c,\bar{c}|V_{ab}|A,B\right >
&=& \int{d\vec{r}_{q_1}d\vec{r}_{\bar{q}_2}d\vec{r}_{c}d\vec{r}_{\bar{c}}}
\psi_{q_1\bar{q}_2c\bar{c}}^+ V_{ab} \psi_{AB}
\nonumber\\
&=& \int d\vec{r}_{q_1\bar{q}_2}d\vec{r}_{c\bar{c}}d\vec{R}_{\rm{total}}
d\vec{r}_{q_1\bar{q}_2,c\bar{c}}
\frac{e^{-i\vec{p}^{~\prime}_{q_1\bar{q}_2}\cdot\vec{r}_{q_1\bar{q}_2}}}
{\sqrt{V}}
\frac{e^{-i\vec{p}^{~\prime}_{c\bar{c}}\cdot\vec{r}_{c\bar{c}}}}{\sqrt{V}}
\frac{e^{-i\vec{p}^{~\prime}_{q_1\bar{q}_2,c\bar{c}}\cdot\vec{r}_{q_1\bar{q}_2,
c\bar{c}}}}{\sqrt{V}}\nonumber\\
&& \times \frac{e^{-i\vec{P}_{\rm f}\cdot\vec{R}_{\rm{total}}}}{\sqrt{V}}
\varphi^{+}_{q_1\bar{q}_2c\bar{c}{\rm color}}
\varphi^{+}_{q_1\bar{q}_2c\bar{c}{\rm flavor}}
\varphi^{+}_{q_1\bar{q}_2c\bar{c}{\rm spin}}
V_{ab}\psi_{A}(\vec{r}_{q_{1}\bar{q}_{2}})\psi_{B}(\vec{r}_{c\bar{c}})
\nonumber\\
&& \times \frac{e^{i\vec{p}_{AB}\cdot \vec{r}_{AB}}}{\sqrt{V}}
\frac{e^{i\vec{P}_{\rm i}\cdot \vec{R}_{\rm{total}}}}{\sqrt{V}}\nonumber\\
&=& (2\pi)^3\delta^3(\vec{P}_{\rm f}-\vec{P}_{\rm i})
\frac{{\cal M}_{ab}}{V^3\sqrt{2E_{A}2E_{B}2E^{\prime}_{q_1}
2E^{\prime}_{\bar{q}_2}
2E^{\prime}_{c}2E^{\prime}_{\bar{c}}}}, 
\end{eqnarray}
where $\vec{R}_{\rm{total}}$ and $\vec{P}_{\rm f}$ are the center-of-mass
coordinate and the total momentum of $q_1$, $\bar{q}_2$, $c$, and $\bar{c}$,
respectively; 
$\vec{r}_{q_1\bar{q}_2,c\bar{c}}$ and 
$\vec{p}_{q_1\bar{q}_2,c\bar{c}}^{~\prime}$ the relative coordinate and the 
relative momentum of the two colored pairs $q_1\bar{q}_2$ and $c\bar{c}$, 
respectively;
$\vec{p}^{~\prime}_{q_1\bar{q}_2}$ the relative momentum of $q_1$ 
and $\bar{q}_2$, and
$\vec{p}^{~\prime}_{c\bar{c}}$ the relative momentum of $c$ and $\bar{c}$;
$\vec{p}_{AB}$ ($\vec{r}_{AB}$, $\vec{P}_{\rm i}$) the relative momentum
(the relative coordinate, the total momentum) of mesons $A$ and $B$; 
$E_A$, $E_B$, $E^{\prime}_{q_1}$, $E^{\prime}_{\bar{q}_2}$, $E^{\prime}_{c}$, 
and $E^{\prime}_{\bar{c}}$ the energies of $A$, $B$, $q_1$, $\bar{q}_2$, $c$, 
and $\bar{c}$, respectively.
${\cal M}_{ab}$ is the transition amplitude corresponding to $V_{ab}$:
\begin{eqnarray}
{\cal M}_{ab} 
&=& \sqrt{2E_A2E_B2E^{\prime}_{q_1}2E^{\prime}_{\bar{q}_2}2E^{\prime}_{c}
2E^{\prime}_{\bar{c}}}
\nonumber\\
&& \times\int d\vec{r}_{q_1\bar{q}_2}d\vec{r}_{c\bar{c}}d\vec{r}_{q_1\bar{q}_2,
c\bar{c}}
e^{-i\vec{p}^{~\prime}_{q_1\bar{q}_2}\cdot\vec{r}_{q_1\bar{q}_2}
-i\vec{p}^{~\prime}_{c\bar{c}}
\cdot\vec{r}_{c\bar{c}}-i\vec{p}^{~\prime}_{q_1\bar{q}_2,c\bar{c}}\cdot
\vec{r}_{q_1\bar{q}_2,c\bar{c}}}
\nonumber\\
&& \times \varphi^{+}_{q_1\bar{q}_2c\bar{c}{\rm color}}
\varphi^{+}_{q_1\bar{q}_2c\bar{c}{\rm flavor}}
\varphi^{+}_{q_1\bar{q}_2c\bar{c}{\rm spin}}V_{ab}(\vec{r}_{ab})\psi_{A}
(\vec{r}_{q_{1}\bar{q}_{2}})
\psi_{B}(\vec{r}_{c\bar{c}})e^{i\vec{p}_{AB}\cdot\vec{r}_{AB}}. 
\label{eq:ta}
\end{eqnarray}

Let $\phi_{A{\rm color}}$ ($\phi_{B{\rm color}}$), $\phi_{A{\rm flavor}}$ 
($\phi_{B{\rm flavor}}$),
$\phi_{A{\rm rel}}$ ($\phi_{B{\rm rel}}$), and $\chi_A$ 
($\chi_B$) stand for the color wave function, the flavor wave 
function, the quark-antiquark relative-motion wave function, and the spin
wave function of meson $A$ ($B$), respectively; denote the total angular 
momentum, the orbital angular
momentum, and the spin of meson $A$ ($B$) by $J_A$ ($J_B$), $L_A$ ($L_B$), and 
$S_A$ ($S_B$), respectively.
$\psi_A$ and $\psi_B$ in Eq. (\ref{eq:wfAB}) are given by
\begin{eqnarray}
\psi_A(\vec{r}_{q_1\bar{q}_2}) &=& \phi_{A{\rm color}}\phi_{A{\rm flavor}}
(\phi_{A{\rm rel}}\chi_A)^{J_A}_{J_{Az}},\\
\psi_B(\vec{r}_{c\bar{c}}) &=& \phi_{B{\rm color}}\phi_{B{\rm flavor}}
(\phi_{B{\rm rel}}\chi_B)^{J_B}_{J_{Bz}},
\end{eqnarray}
where $J_{Az}$ ($J_{Bz}$) is the magnetic projection quantum number of 
$J_A$ ($J_B$), and the symbol $(\cdot \cdot \cdot )_{J_{Az}}^{J_A}$
[$(\cdot \cdot \cdot )_{J_{Bz}}^{J_B}$] indicates the space-spin wave function
of meson $A$ ($B$). The product of $\psi_A$ and $\psi_B$ is
\begin{eqnarray}
\psi_A(\vec{r}_{q_1\bar{q}_2})\psi_B(\vec{r}_{c\bar{c}})=
\phi_{A{\rm color}}\phi_{B{\rm color}}\phi_{A{\rm flavor}}\phi_{B{\rm flavor}}
\sum_{JJ_z}(J_AJ_{Az}J_BJ_{Bz}|JJ_z)\psi^{JJ_z}_{\rm in},
\end{eqnarray}
where $J$ is the total angular momentum of mesons $A$ and $B$, and $J_z$ its 
magnetic projection quantum 
number; $(J_AJ_{Az}J_BJ_{Bz}|JJ_z)$ are the Clebsch-Gordan coefficients. Let 
$L$ $(S)$ and $L_z$ $(S_z)$ denote the total orbital angular momentum 
(total spin) of mesons $A$ and $B$ and its magnetic projection quantum number, 
respectively. $\psi^{JJ_z}_{\rm in}$ comes from the coupling of the 
space-spin states of meson $A$ and of meson $B$:
\begin{eqnarray}
\psi^{JJ_z}_{\rm in} &=& [(\phi_{A{\rm rel}}\chi_A)^{J_A}
(\phi_{B{\rm rel}}\chi_B)^{J_B}]^J_{J_z}\nonumber\\
&=& \sum_{LS}\sqrt{(2J_A+1)(2J_B+1)(2L+1)(2S+1)}
\left\{\begin{matrix}
L_A & S_A & J_A \\
L_B & S_B & J_B \\
L & S & J \\
\end{matrix}\right\}\nonumber\\
&& \times[(\phi_{A{\rm rel}}\phi_{B{\rm rel}})^{L}
(\chi_A\chi_B)^{S}]^{J}_{J_z}\nonumber\\
&=& \sum_{LSL_zS_z}\sqrt{(2J_A+1)(2J_B+1)(2L+1)(2S+1)}
\left\{\begin{matrix}
L_A & S_A & J_A \\
L_B & S_B & J_B \\
L & S & J \\
\end{matrix}\right\}\nonumber\\
&& \times(LL_zSS_z|JJ_z)(\phi_{A{\rm rel}}\phi_{B{\rm rel}})^L_{L_z}
(\chi_A\chi_B)^S_{S_z},
\end{eqnarray}
where the two braces in each of the second and third expressions indicate the 
Wigner $9j$ symbol.

Denote by $\phi_{i\rm{flavor}}$, $\phi_i$, and $\chi_i$ the flavor 
wave function, the space wave function, and the spin wave function of 
constituent quark or antiquark labeled as $i$ $(i=q_1,~\bar{q}_2,~c,~\bar{c})$,
respectively. In fact, $\phi_i$ equals
$e^{i\vec{p}^{~\prime}_i\cdot\vec{r}_i}/\sqrt{V}$ in Eq. (\ref{eq:wf}),
\begin{equation}
\varphi_{q_1\bar{q}_2c\bar{c}{\rm flavor}}=
\phi_{q_1{\rm flavor}}\phi_{\bar{q}_2{\rm flavor}}c\bar{c}, 
\end{equation}
and
\begin{equation}
\varphi_{q_1\bar{q}_2c\bar{c}{\rm spin}}=
\chi_{q_1}\chi_{\bar{q}_2}\chi_c\chi_{\bar{c}}.
\end{equation}
The wave function of $q_1$, $\bar{q}_2$, $c$, and $\bar{c}$ is 
\begin{equation}
\psi_{q_1\bar{q}_2c\bar{c}}=
\phi_{q_1}\phi_{\bar{q}_2}\phi_{c}\phi_{\bar{c}}
\varphi_{q_1\bar{q}_2c\bar{c}{\rm color}}
\varphi_{q_1\bar{q}_2c\bar{c}{\rm flavor}}
\varphi_{q_1\bar{q}_2c\bar{c}{\rm spin}}.
\end{equation}
After the spin states of $q_1$ ($c$) and of $\bar{q}_2$ ($\bar{c}$) 
are coupled to the spin state with the spin
$S_{q_1+\bar{q}_2}^{\prime}$ ($S_{c+\bar{c}}^{\prime}$) and its $z$ component 
$S_{q_1+\bar{q}_2z}^{\prime}$ ($S_{c+\bar{c}z}^{\prime}$),
the spin states with $S_{q_1+\bar{q}_2}^{\prime}$ and
with $S_{c+\bar{c}}^{\prime}$ are coupled to the spin state of $q_1$, 
$\bar{q}_2$, $c$, and $\bar{c}$, which has the spin $S^{\prime}$
and its $z$ component $S^{\prime}_z$. In this way we have
\begin{eqnarray}
\varphi_{q_1\bar{q}_2c\bar{c}{\rm spin}} &=& \sum_{S_{q_1+\bar{q}_2}^{\prime}
S_{q_1+\bar{q}_2z}^{\prime}}
(s_{q_1}s_{q_1z}s_{\bar{q}_2}s_{\bar{q}_2z}|S_{q_1+\bar{q}_2}^{\prime}S_{q_1+
\bar{q}_2z}^{\prime})
(\chi_{q_1}\chi_{\bar{q}_2})^{S_{q_1
+\bar{q}_2}^{\prime}}_{S_{q_1+\bar{q}_2z}^{\prime}}
\nonumber\\
&& \times\sum_{S_{c+\bar{c}}^{\prime}S_{c+\bar{c}z}^{\prime}}
(s_{c}s_{cz}s_{\bar{c}}s_{\bar{c}z}|S_{c+\bar{c}}^{\prime}
S_{c+\bar{c}z}^{\prime})(\chi_c
\chi_{\bar c})^{S_{c+\bar{c}}^{\prime}}_{S_{c+\bar{c}z}^{\prime}}
\nonumber\\
&=& \sum_{S_{q_1+\bar{q}_2}^{\prime}S_{q_1+\bar{q}_2z}^{\prime}
S_{c+\bar{c}}^{\prime}S_{c+\bar{c}z}^{\prime}S^{\prime}S^{\prime}_{z}}
(s_{q_1}s_{q_1z}s_{\bar{q}_2}s_{\bar{q}_2z}|S_{q_1+\bar{q}_2}^{\prime}
S_{q_1+\bar{q}_2z}^{\prime})
\nonumber\\
&& \times(s_{c}s_{cz}s_{\bar{c}}s_{\bar{c}z}|S_{c+\bar{c}}^{\prime}
S_{c+\bar{c}z}^{\prime})
(S_{q_1+\bar{q}_2}^{\prime}S_{q_1+\bar{q}_2z}^{\prime}S_{c+\bar{c}}^{\prime}
S_{c+\bar{c}z}^{\prime}|S^{\prime}S^{\prime}_{z})
\psi^{S^{\prime}S^{\prime}_z}_{\rm final},
\end{eqnarray}
where $s_i$ is the spin of constituent $i$, and $s_{iz}$ its $z$ component;
the spin wave function $\psi^{S^{\prime}S^{\prime}_z}_{\rm final}$ is
\begin{equation}
\psi^{S^{\prime}S^{\prime}_z}_{\rm final}=
[(\chi_{q_1}\chi_{\bar{q}_2})^{S_{q_1+\bar{q}_2}^{\prime}}
(\chi_c\chi_{\bar{c}})^{S_{c+\bar{c}}^{\prime}}]^{S^{\prime}}_{S^{\prime}_z}.
\end{equation}
According to Eq. (\ref{eq:sm}), the transition amplitude for 
$A+B\to q_{1}+\bar{q}_{2}+c+\bar{c}$ is
\begin{equation}
{\cal M}_{\rm fi}={\cal M}_{q_1\bar{c}}+{\cal M}_{\bar{q}_2c}
+{\cal M}_{q_1c}+{\cal M}_{\bar{q}_2\bar{c}} ,
\end{equation}
where ${\cal M}_{q_1\bar{c}}$, ${\cal M}_{\bar{q}_2c}$, ${\cal M}_{q_1c}$, and 
${\cal M}_{\bar{q}_2\bar{c}}$ correspond to $V_{q_1\bar{c}}$, 
$V_{\bar{q}_2c}$, $V_{q_1c}$, and $V_{\bar{q}_2\bar{c}}$, respectively.
Summing over the states of $A$, $B$, $q_1$, $\bar{q}_2$, 
$c$, and $\bar{c}$ gives
\begin{eqnarray}
\sum_{s_{q_1z}s_{\bar{q}_2z}s_{cz}s_{\bar{c}z}}\sum_{J_{Az}J_{Bz}}
|\cal{M}_{\rm fi}|^{\rm 2} &=& 
\sum_{s_{q_1z}s_{\bar{q}_2z}s_{cz}s_{\bar{c}z}JJ_{z}}
|\sqrt{2E_A2E_B2E^{\prime}_{q_1}2E^{\prime}_{\bar{q}_2}2E^{\prime}_{c}
2E^{\prime}_{\bar{c}}} \sum_{ab}
	  \nonumber\\
&& \times 
\int d\vec{r}_{q_1\bar{q}_2}d\vec{r}_{c\bar{c}}d\vec{r}_{q_1\bar{q}_2,c\bar{c}}
e^{-i\vec{p}^{~\prime}_{q_1\bar{q}_2}\cdot\vec{r}_{q_1\bar{q}_2}
-i\vec{p}^{~\prime}_{c\bar{c}}
\cdot\vec{r}_{c\bar{c}}-i\vec{p}^{~\prime}_{q_1\bar{q}_2,c\bar{c}}
\cdot\vec{r}_{q_1\bar{q}_2,c\bar{c}}}
	  \nonumber\\
&& \times 
\varphi^{+}_{q_1\bar{q}_2c\bar{c}{\rm color}}
\varphi^{+}_{q_1\bar{q}_2c\bar{c}{\rm flavor}}
\varphi^{+}_{q_1\bar{q}_2c\bar{c}{\rm spin}}
V_{ab}(\vec{r}_{ab})\psi_{\rm in}^{JJ_z}
	  \nonumber\\
&& \times
\phi_{A{\rm color}}\phi_{B{\rm color}}\phi_{A{\rm flavor}}\phi_{B{\rm flavor}}
e^{i\vec{p}_{AB}\cdot\vec{r}_{AB}}|^{2},
\end{eqnarray}
where $ab$ in $\sum\limits_{ab}$ runs through $q_1\bar{c}$, $\bar{q}_2c$, 
$q_1c$, and $\bar{q}_2\bar{c}$.

The potential $V_{ab}$ consists of the central spin-independent potential
$V_{\rm{si}}$ and the spin-spin interaction $V_{\rm{ss}}$:
\begin{eqnarray}
V_{ab}(\vec {r}_{ab}) = V_{\rm{si}}(\vec {r}_{ab})+V_{\rm{ss}}(\vec {r}_{ab}).
\label{eq:vab}
\end{eqnarray}
Below the critical temperature $T_{\rm c}=0.175$ GeV, the spin-independent
potential is given by
\begin{equation}
V_{\rm {si}}(\vec{r}_{ab})=
-\frac {\vec {\lambda}_a}{2} \cdot \frac {\vec{\lambda}_b}{2}
\xi_1 \left[ 1.3- \left( \frac {T}{T_{\rm c}} \right)^4 \right]
\tanh (\xi_2 r_{ab})
+ \frac {\vec {\lambda}_a}{2} \cdot \frac {\vec {\lambda}_b}{2}
\frac {6\pi}{25} \frac {v(\lambda r_{ab})}{r_{ab}} \exp (-\xi_3 r_{ab}),
\label{eq:vsi}
\end{equation}
where $\xi_1=0. 525$ GeV, $\xi_2=1.5[0.75+0.25 (T/{T_{\rm c}})^{10}]^6$ GeV,
$\xi_3=0. 6$ GeV, and $\lambda=\sqrt{3b_0/16\pi^2 \alpha'}$ in which
$\alpha'=1.04$ GeV$^{-2}$ and $b_{0}=11-\frac{2}{3}N_{f}$ with the
quark flavor number $N_{f}=4$. $\vec {\lambda}_a$ are the
Gell-Mann matrices for the color generators of constituent
$a$. The dimensionless function $v(x)$ is
given by Buchm\"{u}ller and Tye \cite{BT}. 

The spin-spin interaction with relativistic effects \cite{GI} is
\cite{ZX,Xu2002}
\begin{eqnarray}
V_{\rm ss}(\vec {r}_{ab})=
- \frac {\vec {\lambda}_a}{2} \cdot \frac {\vec {\lambda}_b}{2}
\frac {16\pi^2}{25}\frac{d^3}{\pi^{3/2}}\exp(-d^2r^2_{ab}) \frac {\vec {s}_a 
\cdot \vec
{s} _b} {m_am_b}
+ \frac {\vec {\lambda}_a}{2} \cdot \frac {\vec {\lambda}_b}{2}
  \frac {4\pi}{25} \frac {1} {r_{ab}}
\frac {d^2v(\lambda r_{ab})}{dr^2_{ab}} \frac 
{\vec {s}_a \cdot \vec {s}_b}{m_am_b},
\label{eq:vss}
\end{eqnarray}
where $m_a$ is the mass of constituent $a$, and $d$ is given by  
\begin{displaymath}
d^2=d_{1}^2\left[\frac{1}{2}+\frac{1}{2}
\left(\frac{4m_a m_b}{(m_a+m_b)^2}\right)^4\right]
+d_{2}^2\left(\frac{2m_am_b}{m_a+m_b}\right)^2,
\end{displaymath}
where $d_1=0.15$ GeV and $d_2=0.705$.

One-gluon exchange between constituents $a$ and $b$ gives rise to the Fermi 
contact term $- \frac {\vec {\lambda}_a}{2} \cdot \frac {\vec {\lambda}_b}{2}
\frac {16\pi^2}{25} \delta^3 (\vec{r}_{ab}) 
\frac {\vec {s}_a \cdot \vec {s} _b} {m_am_b}$ in the nonrelativistic limit.
The $\delta^3 (\vec{r}_{ab})$ function fixes the positions of the two 
constituents to $\vec{r}_{ab}=0$. However, the constituent positions fluctuate
because each constituent is coupled to a gluon field which has vacuum
polarization. This is similar to the well-known fact that the fluctuation of an
electron position arises from vacuum polarization of its coupled 
electromagnetic field
\cite{BD,Schwabl}. To take into account this relativistic effect, 
$\delta^3 (\vec{r}_{ab})$ is replaced with 
$\frac{d^3}{\pi^{3/2}}\exp(-d^2r^2_{ab})$ so as to arrive at the first term
on the right-hand side of Eq. (19). This is the smearing of the 
one-gluon-exchange spin-spin interaction \cite{GI}. 
The second term on the right-hand side of Eq. (18) comes from one-gluon
exchange plus perturbative one- and two-loop corrections. The second term on 
the right-hand side of Eq. (19) originates from perturbative one- and two-loop
corrections to one-gluon exchange \cite{Xu2002}. The loop corrections are 
another relativistic effect embedded in the spin-independent potential and the
spin-spin interaction. Therefore, the potential $V_{ab}$ given in Eq. (17) is 
a relativized potential. 

The potential at short distances is dominated by the the second term of the 
spin-independent potential and the two terms of the spin-spin interaction. When
the center-of-mass energy of mesons $A$ and $B$ is large, short distances are
reached by constituents, and the three terms with relativistic effects 
make a contribution to the scattering of mesons $A$ and $B$.

The total-spin operator of $A$ and $B$, i.e., of $q_1$, $\bar{q}_2$, $c$, and 
$\bar{c}$ is
\begin{equation}
\vec{s}=\vec{s}_{q_1}+\vec{s}_{\bar{q}_2}+\vec{s}_{c}+\vec{s}_{\bar{c}}.
\end{equation}
It is easily proved that the commutator of $\vec{s}$ and the Hamiltonian that 
includes $V_{ab}$ equals zero. This leads to $S^{\prime}=S$ and 
$S^{\prime}_z=S_z$. We thus get  
\begin{eqnarray}
&& \sum_{s_{q_1z}s_{\bar{q}_2z}s_{cz}s_{\bar{c}z}J_{Az}J_{Bz}}
|\cal{M}_{\rm fi}|^{\rm 2}
	  \nonumber\\
&& =2E_A2E_B2E^{\prime}_{q_1}2E^{\prime}_{\bar{q}_2}2E^{\prime}_{c}
2E^{\prime}_{\bar{c}}
	  \nonumber\\
&& \times
\sum_{LSJL_zS_{q_1+\bar{q}_2}^{\prime}S_{c+\bar{c}}^{\prime}}
(2J_A+1)(2J_B+1)(2S+1)(2J+1)
\left\{\begin{matrix}
L_A & S_A & J_A \\
L_B & S_B & J_B \\
L & S & J \\
\end{matrix}\right\}^2
	  \nonumber\\
&& \times
|\sum_{ab} \int d\vec{r}_{q_1\bar{q}_2}d\vec{r}_{c\bar{c}}
d\vec{r}_{q_1\bar{q}_2,c\bar{c}}
e^{-i\vec{p}^{~\prime}_{q_1\bar{q}_2}\cdot\vec{r}_{q_1\bar{q}_2}
-i\vec{p}^{~\prime}_{c\bar{c}}
\cdot\vec{r}_{c\bar{c}}-i\vec{p}^{~\prime}_{q_1\bar{q}_2,c\bar{c}}
\cdot\vec{r}_{q_1\bar{q}_2,c\bar{c}}+i\vec{p}_{AB} \cdot \vec{r}_{AB}}
	  \nonumber\\
&& \times
\varphi^{+}_{q_1\bar{q}_2c\bar{c}{\rm color}}
\varphi^{+}_{q_1\bar{q}_2c\bar{c}{\rm flavor}}
\psi_{\rm final}^{SS_z+}
V_{ab}(\vec{r}_{ab})\phi_{A{\rm color}}\phi_{B{\rm color}}
\phi_{A{\rm flavor}}\phi_{B{\rm flavor}}
(\phi_{A{\rm rel}}\phi_{B{\rm rel}})^L_{L_z}
	  \nonumber\\
&& \times
(\chi_A\chi_B)^S_{S_z}|^2.
\end{eqnarray}

We take the Fourier transform of the mesonic quark-antiquark relative-motion
wave functions and the potentials:
\begin{eqnarray}
\phi_{A{\rm rel}}(\vec{r}_{q_1\bar{q}_2}) &=& 
\int\frac{d^3p_{q_1\bar{q}_2}}{(2\pi)^3}
\phi_{A{\rm rel}}(\vec{p}_{q_1\bar{q}_2})e^{i\vec{p}_{q_1\bar{q}_2}
\cdot\vec{r}_{q_1\bar{q}_2}},
\label{eq:Arel}\\
\phi_{B{\rm rel}}(\vec{r}_{c\bar{c}}) &=& \int\frac{d^3p_{c\bar{c}}}{(2\pi)^3}
\phi_{B{\rm rel}}(\vec{p}_{c\bar{c}})e^{i\vec{p}_{c\bar{c}}
\cdot\vec{r}_{c\bar{c}}},
\label{eq:Brel}\\
V_{q_1\bar{c}}(\vec{r}_{q_1\bar{c}}) &=& \int\frac{d^3Q}{(2\pi)^3}
V_{q_1\bar{c}}(\vec{Q})e^{i\vec{Q}\cdot\vec{r}_{q_1\bar{c}}},\\
V_{\bar{q}_2c}(\vec{r}_{\bar{q}_2c}) &=& \int\frac{d^3Q}{(2\pi)^3}
V_{\bar{q}_2c}(\vec{Q})e^{i\vec{Q}\cdot\vec{r}_{\bar{q}_2c}},\\
V_{q_1c}(\vec{r}_{q_1c}) &=& \int\frac{d^3Q}{(2\pi)^3}
V_{q_1c}(\vec{Q})e^{i\vec{Q}\cdot\vec{r}_{q_1c}},\\
V_{\bar{q}_2\bar{c}}(\vec{r}_{\bar{q}_2\bar{c}}) &=& \int\frac{d^3Q}{(2\pi)^3}
V_{\bar{q}_2\bar{c}}(\vec{Q})e^{i\vec{Q}\cdot\vec{r}_{\bar{q}_2\bar{c}}}.
\end{eqnarray}
The quark-antiquark relative-motion wave functions in momentum space, 
$\phi_{A{\rm rel}}$ $(\vec{p}_{q_1\bar{q}_2})$ and $\phi_{B{\rm rel}}$ 
$(\vec{p}_{c\bar{c}})$,
satisfy $\int\frac{d^3p_{q_1\bar{q}_2}}{(2\pi)^3}\phi^+_{A{\rm rel}}
(\vec{p}_{q_1\bar{q}_2})\phi_{A{\rm rel}}(\vec{p}_{q_1\bar{q}_2})
=\int\frac{d^3p_{c\bar{c}}}{(2\pi)^3}\phi^+_{B{\rm rel}}(\vec{p}_{c\bar{c}})
\phi_{B{\rm rel}}(\vec{p}_{c\bar{c}})=1$. We finally arrive at
\begin{eqnarray}
&& \sum_{s_{q_1z}s_{\bar{q}_2z}s_{cz}s_{\bar{c}z}J_{Az}J_{Bz}}
|\cal{M}_{\rm fi}|^{\rm 2}
	  \nonumber\\
&& =2E_A2E_B2E^{\prime}_{q_1}2E^{\prime}_{\bar{q}_2}2E^{\prime}_{c}
2E^{\prime}_{\bar{c}}
	  \nonumber\\
&& \times\sum_{LSJL_zS_{q_1+\bar{q}_2}^{\prime}S_{c+\bar{c}}^{\prime}}
(2J_A+1)(2J_B+1)(2S+1)(2J+1)
\left\{\begin{matrix}
L_A & S_A & J_A \\
L_B & S_B & J_B \\
L & S & J \\
\end{matrix}\right\}^2
	  \nonumber\\
&& \times|\varphi^{+}_{q_1\bar{q}_2c\bar{c}{\rm color}}\varphi^{+}_{q_1
\bar{q}_2c\bar{c}{\rm flavor}}
\psi_{\rm final}^{SS_z+}
	  \nonumber\\
&&\times  \left\{ V_{q_1\bar{c}}(\vec{Q})
{\left[\phi_{A{\rm rel}}(\vec{p}^{~\prime}_{q_1\bar{q}_2}
-\frac{m_{\bar{q}_2}}{m_{q_1}+m_{\bar{q}_2}}\vec{Q})
\phi_{B{\rm rel}}(\vec{p}^{~\prime}_{c\bar{c}}
-\frac{m_c}{m_c+m_{\bar{c}}}\vec{Q})
\right]}^L_{L_z}\right.
     \nonumber\\
&&+V_{\bar{q}_2c}(\vec{Q})
{\left[\phi_{A{\rm rel}}(\vec{p}^{~\prime}_{q_1\bar{q}_2}
+\frac{m_{q_1}}{m_{q_1}+m_{\bar{q}_2}}\vec{Q})
\phi_{B{\rm rel}}(\vec{p}^{~\prime}_{c\bar{c}}
+\frac{m_{\bar{c}}}{m_c+m_{\bar{c}}}\vec{Q})
\right]}^L_{L_z}
     \nonumber\\
&&+V_{q_1c}(\vec{Q})
{\left[\phi_{A{\rm rel}}(\vec{p}^{~\prime}_{q_1\bar{q}_2}
-\frac{m_{\bar{q}_2}}{m_{q_1}+m_{\bar{q}_2}}\vec{Q})
\phi_{B{\rm rel}}(\vec{p}^{~\prime}_{c\bar{c}}
+\frac{m_{\bar{c}}}{m_c+m_{\bar{c}}}\vec{Q})
\right]}^L_{L_z}
     \nonumber\\
&&\left.+V_{\bar{q}_2\bar{c}}(\vec{Q})
{\left[\phi_{A{\rm rel}}(\vec{p}^{~\prime}_{q_1\bar{q}_2}
+\frac{m_{q_1}}{m_{q_1}+m_{\bar{q}_2}}\vec{Q})
\phi_{B{\rm rel}}(\vec{p}^{~\prime}_{c\bar{c}}
-\frac{m_c}{m_c+m_{\bar{c}}}\vec{Q})
\right]}^L_{L_z}\right\}
	  \nonumber\\
&& \times(\chi_A\chi_B)^S_{S_z}\phi_{A{\rm flavor}}
\phi_{B{\rm flavor}}\phi_{A{\rm color}}\phi_{B{\rm color}}|^2,
\label{eq:qarf}
\end{eqnarray}
where $\vec{Q}$ is the gluon momentum.

Let $d\sigma_{\rm free}$ represent the differential cross section 
corresponding to the factor 
$\psi_{\rm final}^{SS_z+}\sum_{ab}$
$V_{ab}(\phi_{A{\rm rel}}\phi_{B{\rm rel}})^{L}_{L_z}
(\chi_A\chi_B)^S_{S_z}$ in Eq. (28),
\begin{eqnarray}
d\sigma_{\rm free}&=&
\frac{(2\pi)^4}{4\sqrt{(P_A\cdot P_B)^2-m_A^2m_B^2}}
\frac{d^3p^{\prime}_{q_1}}{(2\pi)^32E^{\prime}_{q_1}}
\frac{d^3p^{\prime}_{\bar{q}_2}}{(2\pi)^32E^{\prime}_{\bar{q}_2}}
\frac{d^3p^{\prime}_{c}}{(2\pi)^32E^{\prime}_{c}}
\frac{d^3p^{\prime}_{\bar{c}}}{(2\pi)^32E^{\prime}_{\bar{c}}}
	  \nonumber\\
& \times &
\delta^4(P_A+P_B-p^{\prime}_{q_1}-p^{\prime}_{\bar{q}_2}-p^{\prime}_{c}
-p^{\prime}_{\bar{c}})
	  \nonumber\\
& \times & 
2E_A2E_B2E^{\prime}_{q_1}2E^{\prime}_{\bar{q}_2}2E^{\prime}_{c}
|\varphi^{+}_{q_1\bar{q}_2c\bar{c}{\rm color}}\varphi^{+}_{q_1
\bar{q}_2c\bar{c}{\rm flavor}}
\psi_{\rm final}^{SS_z+}
	  \nonumber\\
& \times & \left\{ V_{q_1\bar{c}}(\vec{Q})
{\left[\phi_{A{\rm rel}}(\vec{p}^{~\prime}_{q_1\bar{q}_2}
-\frac{m_{\bar{q}_2}}{m_{q_1}+m_{\bar{q}_2}}\vec{Q})
\phi_{B{\rm rel}}(\vec{p}^{~\prime}_{c\bar{c}}
-\frac{m_c}{m_c+m_{\bar{c}}}\vec{Q})
\right]}^L_{L_z}\right.
     \nonumber\\
& + & V_{\bar{q}_2c}(\vec{Q})
{\left[\phi_{A{\rm rel}}(\vec{p}^{~\prime}_{q_1\bar{q}_2}
+\frac{m_{q_1}}{m_{q_1}+m_{\bar{q}_2}}\vec{Q})
\phi_{B{\rm rel}}(\vec{p}^{~\prime}_{c\bar{c}}
+\frac{m_{\bar{c}}}{m_c+m_{\bar{c}}}\vec{Q})
\right]}^L_{L_z}
     \nonumber\\
& + & V_{q_1c}(\vec{Q})
{\left[\phi_{A{\rm rel}}(\vec{p}^{~\prime}_{q_1\bar{q}_2}
-\frac{m_{\bar{q}_2}}{m_{q_1}+m_{\bar{q}_2}}\vec{Q})
\phi_{B{\rm rel}}(\vec{p}^{~\prime}_{c\bar{c}}
+\frac{m_{\bar{c}}}{m_c+m_{\bar{c}}}\vec{Q})
\right]}^L_{L_z}
     \nonumber\\
& + & \left. V_{\bar{q}_2\bar{c}}(\vec{Q})
{\left[\phi_{A{\rm rel}}(\vec{p}^{~\prime}_{q_1\bar{q}_2}
+\frac{m_{q_1}}{m_{q_1}+m_{\bar{q}_2}}\vec{Q})
\phi_{B{\rm rel}}(\vec{p}^{~\prime}_{c\bar{c}}
-\frac{m_c}{m_c+m_{\bar{c}}}\vec{Q})
\right]}^L_{L_z}\right\}
	  \nonumber\\
& \times & (\chi_A\chi_B)^S_{S_z}\phi_{A{\rm flavor}}
\phi_{B{\rm flavor}}\phi_{A{\rm color}}\phi_{B{\rm color}}|^2,
\end{eqnarray}
where $m_A$ ($m_B$) is the mass of meson $A$ ($B$); $P_A=(E_A,~\vec{P}_A)$, 
$P_B=(E_B,~\vec{P}_B)$, $s=(P_A+P_B)^2$, and 
$p^{\prime}_i=(E^{\prime}_i,~\vec{p}^{~\prime}_i)$
with $i=q_1,~\bar{q}_2,~c$, and $\bar{c}$.
$d\sigma_{\rm free}$ depends on $L_A$, $S_A$, $L_B$, $S_B$, $L$, $S$, $L_z$,
$S_{q_1+\bar{q}_2}^{\prime}$, and $S_{c+\bar{c}}^{\prime}$. The unpolarized 
differential cross section for $A+B\to q_{1}+\bar{q}_{2}+c+\bar{c}$ is thus
\begin{eqnarray}
d\sigma_{\rm free}^{\rm unpol}(\sqrt{s},T) &=&
\frac{(2\pi)^4}{4\sqrt{(P_A\cdot P_B)^2-m_A^2m_B^2}}
\frac{d^3p^{\prime}_{q_1}}{(2\pi)^32E^{\prime}_{q_1}}
\frac{d^3p^{\prime}_{\bar{q}_2}}{(2\pi)^32E^{\prime}_{\bar{q}_2}}
\frac{d^3p^{\prime}_{c}}{(2\pi)^32E^{\prime}_{c}}
\frac{d^3p^{\prime}_{\bar{c}}}{(2\pi)^32E^{\prime}_{\bar{c}}}
	  \nonumber\\
& \times &
\delta^4(P_A+P_B-p^{\prime}_{q_1}-p^{\prime}_{\bar{q}_2}-p^{\prime}_{c}
-p^{\prime}_{\bar{c}})
\sum_{s_{q_1z}s_{\bar{q}_2z}s_{cz}s_{\bar{c}z}J_{Az}J_{Bz}}
|\cal{M}_{\rm fi}|^{\rm 2}
	  \nonumber\\
&=& \sum_{LSJL_zS_{q_1+\bar{q}_2}^{\prime}S_{c+\bar{c}}^{\prime}}
(2S+1)(2J+1)
\left\{\begin{matrix}
L_A & S_A & J_A \\
L_B & S_B & J_B \\
L & S & J \\
\end{matrix}\right\}^2
d\sigma_{\rm free}.
\label{eq:udcs1}
\end{eqnarray}

Now we give the cross section for 
$A+B\to q_{1}+\bar{q}_{2}+c+\bar{c}\to H_c+X$, which includes the 
fragmentation process $c\to H_c$. Denote by $z$ the fraction of energy passed 
on from quark $c$ to hadron $H_c$. The fragmentation function 
$D^{H_c}_c(z,{\mu}^2)$ at the factorization scale $\mu$ indicates that 
$D^{H_c}_c(z,{\mu}^2)dz$ is the number of hadron $H_c$ produced at $z$ and 
within $dz$. Consequently, the unpolarized differential cross section for 
$A+B\to q_{1}+\bar{q}_{2}+c+\bar{c}\to H_c+X$ is
\begin{eqnarray}
d\sigma^{\rm unpol}(\sqrt{s},T) & = &
d\sigma_{\rm free}^{\rm unpol}(\sqrt{s},T) D^{H_c}_c(z,{\mu}^2)dz
  \nonumber \\
& = & \sum_{LSJL_zS_{q_1+\bar{q}_2}^{\prime}
S_{c+\bar{c}}^{\prime}}
(2S+1)(2J+1)
\left\{\begin{matrix}
L_A & S_A & J_A \\
L_B & S_B & J_B \\
L & S & J \\
\end{matrix}\right\}^2
d\sigma_{\rm free}D^{H_c}_c(z,{\mu}^2)dz .
  \nonumber \\
\end{eqnarray}

The present work involves the three cases: $L_A=0$, $L_B=0$; $L_A=0$, 
$L_B\neq0$, 
$S_A=0$; $L_A=0$, $L_B=1$, $S_A=1$, $S_B=1$. Values of the Wigner $9j$ symbol 
in these cases reduce the unpolarized differential cross section to
\begin{equation}
d\sigma^{\rm{unpol}}(\sqrt{s},T) = \frac{1}{(2S_A+1)(2S_B+1)(2L_B+1)}
\sum\limits_{L_{Bz}SS_{q_1+\bar{q}_2}^{\prime}S_{c+\bar{c}}^{\prime}} 
(2S+1)d\sigma_{\rm free}D^{H_c}_c(z,{\mu}^2)dz.
\label{eq:udcs2}
\end{equation}
The unpolarized cross section for $A+B\to q_{1}+\bar{q}_{2}+c+\bar{c}\to 
H_c+X$ is
\begin{eqnarray}
\sigma^{\rm{unpol}}(\sqrt{s},T)=\frac{1}{(2S_A+1)(2S_B+1)(2L_B+1)}\sum
\limits_{L_{Bz}SS_{q_1+\bar{q}_2}^{\prime}S_{c+\bar{c}}^{\prime}}
(2S+1)\sigma(\sqrt{s},T),
\label{eq:uc}
\end{eqnarray}
with
\begin{equation}
\sigma(\sqrt{s},T) = \int d\sigma_{\rm free}D_{c}^{H_{c}}(z,\mu^2)dz.
\end{equation}
The cross section depends on temperature and the center-of-mass energy 
$\sqrt{s}$ of mesons $A$ and $B$.

\vspace{0.5cm}
\leftline{\bf 3. Numerical results and discussions}
\vspace{0.5cm}

For large-momentum charmonia we consider the following charmonium dissociation
reactions:
\begin{displaymath}
\pi+J/\psi\to H_c+X,~\pi+\psi'\to H_c+X,~\pi+\chi_c\to H_c+X, 
\end{displaymath}
\begin{displaymath}
\rho+J/\psi\to H_c+X,~\rho+\psi'\to H_c+X,~\rho+\chi_c\to H_c+X, 
\end{displaymath}
\begin{displaymath}
K+J/\psi\to H_c+X,~K+\psi'\to H_c+X,~K+\chi_c\to H_c+X, 
\end{displaymath}
\begin{displaymath}
K^{*}+J/\psi\to H_c+X,~K^{*}+\psi'\to H_c+X,~K^{*}+\chi_c\to H_c+X, 
\end{displaymath}
where $H_c$ is $D^+$, $D^0$, $D_s^+$, or $D^{*+}$.
We solve the Schr\"{o}dinger equation with the potential given in Eq. 
(\ref{eq:vab}) to obtain 
$\phi_{A\rm rel} (\vec{r}_{q_1\bar{q}_2})$, 
$\phi_{B\rm rel}(\vec{r}_{c\bar c})$, and temperature-dependent meson masses 
where the up-quark mass, the strange-quark mass, and the charm-quark mass are
0.32, 0.5, and 1.51 GeV,
respectively. The momentum-space wave functions 
[$\phi_{A\rm rel} (\vec{p}_{q_1\bar{q}_2})$ and 
$\phi_{B\rm rel}(\vec{p}_{c\bar c})$] appearing in
Eqs. (\ref{eq:Arel}) and (\ref{eq:Brel}) are used in 
Eq. (29) to calculate $d\sigma_{\rm free}$. 

According to Eqs. (\ref{eq:vab})-(\ref{eq:vss}) and (29),
we need to calculate the color matrix elements 
$\varphi^+_{q_1\bar{q}_2c\bar{c}{\rm color}}
\frac{\vec{\lambda}_a}{2}\cdot\frac{\vec{\lambda}_b}{2}\phi_{A{\rm color}}
\phi_{B{\rm color}}$. $\varphi_{q_1\bar{q}_2c\bar{c}{\rm color}}$ is derived 
in the appendix. Corresponding to the potentials
$V_{q_1\bar{c}}$, $V_{\bar{q}_2c}$, $V_{q_1c}$, and $V_{\bar{q}_2\bar{c}}$, 
the color matrix elements are
\begin{eqnarray}
\varphi^+_{q_1\bar{q}_2c\bar{c}{\rm color}}\frac{\vec{\lambda}_{q_1}}{2}
\cdot\frac{\vec{\lambda}_{\bar{c}}}{2}\phi_{A{\rm color}}\phi_{B{\rm color}} 
&=& -\frac{\sqrt{6}}{9},\\
\varphi^+_{q_1\bar{q}_2c\bar{c}{\rm color}}\frac{\vec{\lambda}_{\bar{q}_2}}{2}
\cdot\frac{\vec{\lambda}_{c}}{2}\phi_{A{\rm color}}\phi_{B{\rm color}} &=& 
-\frac{\sqrt{6}}{9},\\
\varphi^+_{q_1\bar{q}_2c\bar{c}{\rm color}}\frac{\vec{\lambda}_{q_1}}{2}
\cdot\frac{\vec{\lambda}_{c}}{2}\phi_{A{\rm color}}\phi_{B{\rm color}} &=& 
\frac{\sqrt{6}}{9},\\
\varphi^+_{q_1\bar{q}_2c\bar{c}{\rm color}}\frac{\vec{\lambda}_{\bar{q}_2}}{2}
\cdot\frac{\vec{\lambda}_{\bar{c}}}{2}\phi_{A{\rm color}}\phi_{B{\rm color}} 
&=& \frac{\sqrt{6}}{9}.
\end{eqnarray}
The flavor matrix elements
$\varphi^+_{q_1\bar{q}_2c\bar{c}{\rm flavor}}
\phi_{A{\rm flavor}}\phi_{B{\rm flavor}}$ in Eq. (29) are 1.

According to Eqs. (\ref{eq:vab})-(\ref{eq:vss}) and 
(29), we calculate 
$\psi^{SS_z+}_{\rm final}(\chi_A\chi_B)^S_{S_z}$ for the 
central spin-independent potential and 
$\psi^{SS_z+}_{\rm final}\vec{s}_a\cdot\vec{s}_b(\chi_A\chi_B)^S_{S_z}$
for the spin-spin interaction. These spin matrix elements are listed in 
Table 1. They are independent of $S_z$, and $d\sigma_{\rm free}$ is thus 
independent of $S_z$.

The charm-quark fragmentation functions used in Eq. (34) are 
solutions of the Dokshitzer-Gribov-Lipatov-Altarelli-Paris (DGLAP) evolution
equations with $\mu=\sqrt s$ \cite{KK,KKKS}. 
The starting point for the DGLAP evolution in 
$\mu$ is taken to be the charm-quark mass.
Unpolarized cross sections for charmonium dissociation in collisions with
$\pi$, $\rho$, $K$, and $K^*$ mesons are 
calculated with Eq. (33). For the convenient
use of the unpolarized cross sections, they are parametrized as
\begin{eqnarray}
\sigma^{\rm unpol}(\sqrt {s},T)
&=&a_1 \left( \frac {\sqrt {s} -\sqrt {s_0}} {b_1} \right)^{c_1}
\exp \left[ c_1 \left( 1-\frac {\sqrt {s} -\sqrt {s_0}} {b_1} \right) \right]
\nonumber \\
&&+ a_2 \left( \frac {\sqrt {s} -\sqrt {s_0}} {b_2} \right)^{c_2}
\exp \left[ c_2 \left( 1-\frac {\sqrt {s} -\sqrt {s_0}} {b_2} \right) \right],
\label{eq:csp}
\end{eqnarray}
where $\sqrt{s_0}$ is the threshold energy and equals the sum of
the $H_c$ mass, the $\bar{D}$ mass, the $q_1$ mass, and the $\bar{q}_2$ mass.
The values of the parameters, $a_1$, $b_1$, $c_1$, $a_2$, $b_2$,
and $c_2$, are listed in Tables 2-10. 
In these tables, $d_0$ is the separation between the peak's location on
the $\sqrt{s}$-axis and the threshold energy, and $\sqrt{s_{\rm
z}}$ is the square root of the Mandelstam variable at which the
cross section is 1/100 of the peak cross section that is obtained from the 
parametrization. We note that the parametrization of a reaction at a given 
temperature is valid in the $\sqrt s$ region where the cross-section curve for
the reaction is displayed below.

Cross sections for the pion-charmonium reactions were obtained with 
an early version of FORTRAN code in Ref.\cite{JXH}. After an error
is removed, a new version is used to calculate pion-charmonium dissociation
cross sections, which are smaller than those shown in Ref. \cite{JXH}.
We do not plot the cross sections, but 
list values of $a_1$, $b_1$, $c_1$, $a_2$, $b_2$, $c_2$, $d_0$, and 
$\sqrt{s_z}$ in Tables 11-13.  

In Figs. 1-9 we plot unpolarized cross sections for the following reactions:
\begin{displaymath}
\rho J/\psi \to D^+X + D^0X + D_s^+X + D^{*+}X,
\end{displaymath}
\begin{displaymath}
\rho \psi^\prime \to D^+X + D^0X + D_s^+X + D^{*+}X,
\end{displaymath}
\begin{displaymath}
\rho \chi_c \to D^+X + D^0X + D_s^+X + D^{*+}X,
\end{displaymath}
\begin{displaymath}
K J/\psi \to D^+X + D^0X + D_s^+X + D^{*+}X,
\end{displaymath}
\begin{displaymath}
K \psi^\prime \to D^+X + D^0X + D_s^+X + D^{*+}X,
\end{displaymath}
\begin{displaymath}
K \chi_c \to D^+X + D^0X + D_s^+X + D^{*+}X,
\end{displaymath}
\begin{displaymath}
K^\ast J/\psi \to D^+X + D^0X + D_s^+X + D^{*+}X,
\end{displaymath}
\begin{displaymath}
K^\ast \psi^\prime \to D^+X + D^0X + D_s^+X + D^{*+}X,
\end{displaymath}
\begin{displaymath}
K^\ast \chi_c \to D^+X + D^0X + D_s^+X + D^{*+}X.
\end{displaymath}
Since mesons $A$ and $B$ are broken up in the reaction 
$A+B \to q_1+\bar{q}_2+c+\bar c$, the produced constituents $q_1$, 
$\bar{q}_2$, $c$, and $\bar c$ are described by plane waves. The first term 
on the right-hand side of Eq. (18) stands for the confining potential.
In the confinement regime the mesonic quark-antiquark relative-motion wave
functions are mainly determined by the confining potential, and are 
nonperturbative.
The perturbative part of the mesonic quark-antiquark relative-motion wave
functions is mainly determined by the second term on the right-hand side of
Eq. (18) and the spin-spin
interaction. At low energies near threshold,
the nonperturbative part of the wave functions of mesons $A$ and $B$ overlap,
and it is very difficult for the collision of mesons $A$ and $B$ to break 
them
to produce plane waves of $q_1$, $\bar{q}_2$, $c$, and $\bar c$. Therefore,
the unpolarized cross sections near threshold are negligible. With increasing
$\sqrt s$, more and more the perturbative part of the meson wave functions is
probed, the production of plane waves gradually increases, and the cross 
sections increase.

We present the unpolarized cross sections in the $\sqrt s$ region 
that are generally accessed by collisions between light mesons
in hadronic matter and the three charmonia ($J/\psi$, $\psi'$, and $\chi_c$). 
Cross sections for some reactions at some temperatures reach maximum values
around $\sqrt s=$ 11 GeV. Examples are the cross sections for 
$\rho J/\psi \to D^{*+} X$ at $T/T_{\rm c}=0.95$ in Fig. 10, 
$\rho \psi' \to D_s^+ X$ at $T/T_{\rm c}=0.65$, 0.75, 0.85, 0.9, and 0.95
in Fig. 11, and $K^* \chi_c \to D^0 X$ at $T/T_{\rm c}=0.85$ and 0.9
in Fig. 12. Therefore, we use $\sqrt{s}=11$ GeV in discussions in the 
next two paragraphs.

At large $\sqrt s$ values, the dependence of the cross sections on
temperature is obvious. For example, at $\sqrt s =11$ GeV the cross sections
for $\rho J/\psi$ reactions decrease
with increasing temperature, but the cross sections for 
$\rho \psi^\prime$, $\rho \chi_c$, $K$+charmonium, and $K^*$+charmonium
reactions increase first and then decrease.
Denote by $\sigma_{11\rm min}$ ($\sigma_{11\rm max}$) the smallest (largest) 
cross section among the six cross sections
corresponding to $T/T_{\rm c}$=0, 0.65, 0.75, 0.85, 0.9, and 0.95 at 
$\sqrt {s}$=11 GeV. 
Since $\sigma_{11\rm min}$ is the cross section at $T/T_{\rm c}$=0.95
as seen in Figs. 1-9, it is used as a benchmark
to see the temperature dependence of cross sections for all reactions.
The ratio of $\sigma_{11\rm max}$ to $\sigma_{11\rm min}$ may
represent the variation of the cross section with respect to
temperature. For example, from the dashed curve and the dot-dot-dashed curve
in Fig. 8, $\sigma_{11\rm max} / \sigma_{11\rm min}$ equals 7.46
for $K^* \psi^\prime \to D^+ X + D^0 X + D_s^+ X + D^{*+} X$. The value
indicates that the cross sections for
$K^* \psi^\prime$ reactions change rapidly with increasing temperature.
In order of decrease of $\sigma_{11\rm max} / \sigma_{11\rm min}$,
we list the reactions that produce $D^+ X$, $D^0 X$, $D_s^+ X$, and 
$D^{*+} X$: $K^* \psi'$, $\rho \psi'$, $K^* \chi_c$, $\rho \chi_c$, $K \psi'$,
$K^* J/\psi$, $\rho J/\psi $, $\pi \psi'$, $K \chi_c$, $\pi \chi_c$, 
$K J/\psi$, and $\pi J/\psi$.

A low-energy reaction between a light meson and a charmonium produces two
charmed mesons. While $\sqrt s$ increases from threshold, the cross section
for every endothermic reaction rises from zero, arrives at a maximum value, 
and decreases, but the cross section
for every exothermic reaction decreases rapidly from infinity and then may
increase, reaching a maximum and decreasing.
Peak cross sections of $J/\psi$ dissociation in collisions
with $\pi$, $\rho$, $K$, and $K^*$ mesons are collected from some references,
and are listed in Table 14. In the last row of the table, we show cross 
sections at $\sqrt {s}=11$ GeV and $T=0$ GeV for 
$\pi J/\psi \to D^+X + D^0X + D_s^+X + D^{*+}X + D^{*0}X + D_s^{*+}X$, 
$\rho J/\psi \to D^+X + D^0X + D_s^+X + D^{*+}X + D^{*0}X + D_s^{*+}X$, 
$K J/\psi \to D^+X + D^0X + D_s^+X + D^{*+}X + D^{*0}X + D_s^{*+}X$, and 
$K^{*} J/\psi \to D^+X + D^0X + D_s^+X + D^{*+}X + D^{*0}X + D_s^{*+}X$,
which are obtained in the present work.
Since $c\to D^{*0}$ and $c\to D_s^{*+}$ fragmentation functions are
unknown, cross sections for $\pi J/\psi \to D^{*0}X$, 
$\pi J/\psi \to D_s^{*+}X$, $\rho J/\psi \to D^{*0}X$, 
$\rho J/\psi \to D_s^{*+}X$, $K J/\psi \to D^{*0}X$, $K J/\psi \to D_s^{*+}X$,
$K^* J/\psi \to D^{*0}X$, and $K^* J/\psi \to D_s^{*+}X$ are not calculated.
In order to estimate the eight cross sections at $\sqrt {s}=11$ GeV and $T=0$
GeV, it is assumed that the ratio of the cross section for inclusive $D^{*0}$
($D_s^{*+}$) production to the one for inclusive $D^{*+}$ production equals the
ratio of the one for inclusive $D^0$ ($D_s^+$) production to the one for 
inclusive $D^+$ production, that is,
\begin{displaymath}
\frac{\sigma_{\pi J/\psi \to D^{*0}X}^{\rm unpol}}
{\sigma_{\pi J/\psi \to D^{*+}X}^{\rm unpol}}=
\frac{\sigma_{\pi J/\psi \to D^0X}^{\rm unpol}}
{\sigma_{\pi J/\psi \to D^+X}^{\rm unpol}},
\end{displaymath}
\begin{displaymath}
\frac{\sigma_{\pi J/\psi \to D_s^{*+}X}^{\rm unpol}}
{\sigma_{\pi J/\psi \to D^{*+}X}^{\rm unpol}}=
\frac{\sigma_{\pi J/\psi \to D_s^+X}^{\rm unpol}}
{\sigma_{\pi J/\psi \to D^+X}^{\rm unpol}},
\end{displaymath}
\begin{displaymath}
\frac{\sigma_{\rho J/\psi \to D^{*0}X}^{\rm unpol}}
{\sigma_{\rho J/\psi \to D^{*+}X}^{\rm unpol}}=
\frac{\sigma_{\rho J/\psi \to D^0X}^{\rm unpol}}
{\sigma_{\rho J/\psi \to D^+X}^{\rm unpol}},
\end{displaymath}
and so on. 
It is shown from the table that the cross sections at high $\sqrt s$ in the
present work are
comparable to those peak cross sections at low $\sqrt s$.

\vspace{0.5cm}
\leftline{\bf 4. Summary}
\vspace{0.5cm}

We have obtained temperature-dependent cross sections for dissociation of
large-momentum charmonia in collisions with $\pi$, $\rho$, $K$, and $K^{*}$
mesons in a mechanism where the collision between a light meson and a 
charmonium produces two quarks and two antiquarks first, then the charm quark
fragments into charmed mesons, and the other three constituents combine with
quarks and antiquarks created from vacuum to form two or more mesons.
According to the mechanism, we have derived the transition amplitude for
$A+B \to q_1+\bar{q}_2+c+\bar{c}$ from the wave functions of mesons $A$ and $B$
and of $q_1$, $\bar{q}_2$, $c$, and $\bar c$. An expression for the absolute
square of the transition amplitude is derived from the mesonic quark-antiquark 
relative-motion wave functions and the spin wave functions. The color wave 
function of $q_1$, $\bar{q}_2$, $c$, and $\bar c$ is derived in group
representation theory. With the charm-quark fragmentation functions, the 
unpolarized cross section for $A+B \to q_1+\bar{q}_2+c+\bar{c} \to H_c+X$
is derived. We have calculated the color, flavor, and spin matrix elements.
The mesonic quark-antiquark relative-motion wave functions result from 
the Schr\"odinger equation with the temperature-dependent quark potential. 
We have provided parametrizations of the numerical unpolarized cross sections.

The unpolarized cross sections for inclusive $D^+$, $D^0$, $D_s^+$, and 
$D^{*+}$ production increase with increasing center-of-mass energy of the
colliding light meson and charmonium, depending on the contributions of
the nonperturbative part and the perturbative part of
the quark-antiquark relative-motion wave functions of the colliding mesons.
At high $\sqrt s$ the change of the unpolarized cross sections with
increasing temperature is obvious. Compared to the peak cross sections obtained
in the effective meson approach and in the quark-interchange approach at low
$\sqrt s$, a similar scale of cross sections at high $\sqrt s$ has been
obtained.

\vspace{0.5cm}
\leftline{\bf Acknowledgements}
\vspace{0.5cm}
This work was supported by the National Natural Science Foundation of China 
under Grant No. 11175111.

\vspace{0.5cm}
\leftline{\bf Appendix: The color wave function of $q_1$, $\bar{q}_2$, $c$, 
and $\bar{c}$}
\vspace{0.5cm}

Based on Ref. \cite{Chen}, 
we provide a way to construct the color wave function of $q_1$, $\bar{q}_2$, 
$c$, and $\bar{c}$ that are produced in a collision of mesons
$A$ and $B$. The three color  wave functions of a quark (an antiquark) are 
denoted by $r$, $g$, and $b$ 
($\bar{r}$, $\bar{g}$, and $\bar{b}$), respectively. To apply a permutation
group to obtain the color wave function 
$\varphi_{q_1\bar{q}_2c\bar{c}{\rm color}}$, we use 1, 2, 3, and 4 to label 
$q_1$, $\bar{q}_2$, $c$, and $\bar{c}$, respectively.
The numbers 1 and 2 form a $S_2$ group, and the numbers 3 and 4 
form another $S_2$ group. The direct product of the two permutation groups 
is the group that has the following four elements:
\begin{equation*}
R_1=
\begin{pmatrix}
1 & 2 & 3 & 4\\
1 & 2 & 3 & 4
\end{pmatrix},~
R_2=
\begin{pmatrix}
1 & 2 & 3 & 4\\
2 & 1 & 3 & 4
\end{pmatrix},~
R_3=
\begin{pmatrix}
1 & 2 & 3 & 4\\
1 & 2 & 4 & 3
\end{pmatrix},~
R_4=
\begin{pmatrix}
1 & 2 & 3 & 4\\
2 & 1 & 4 & 3
\end{pmatrix}.
\end{equation*}
Each group element forms a class, and the $S_2 \times S_2$ group has four 
classes. The class 
operator is defined as the sum of all group elements in the class. 
The group has the following four class operators:
\begin{equation}
C_1=R_1,~C_2=R_2,~C_3=R_3,~C_4=R_4.
\end{equation}
In the class space, the class operators give
\begin{equation}
C_iC_j=\sum^4_{k=1}D_{kj}(C_i)C_k ,
\end{equation}
where $D_{kj}(C_i)$ form a matrix that defines a representation of $C_i$ as
\[
D(C_1)=
\begin{pmatrix}
1 & 0 & 0 & 0\\
0 & 1 & 0 & 0\\
0 & 0 & 1 & 0\\
0 & 0 & 0 & 1
\end{pmatrix},~
D(C_2)=
\begin{pmatrix}
0 & 1 & 0 & 0\\
1 & 0 & 0 & 0\\
0 & 0 & 0 & 1\\
0 & 0 & 1 & 0
\end{pmatrix},
\]
\[
D(C_3)=
\begin{pmatrix}
0 & 0 & 1 & 0\\
0 & 0 & 0 & 1\\
1 & 0 & 0 & 0\\
0 & 1 & 0 & 0
\end{pmatrix},~
D(C_4)=
\begin{pmatrix}
0 & 0 & 0 & 1\\
0 & 0 & 1 & 0\\
0 & 1 & 0 & 0\\
1 & 0 & 0 & 0
\end{pmatrix}.
\]
Let $Q$ and $\lambda_i$ individually stand for the eigenvector and the
eigenvalue of $C_i$ in the class space, and $Q$ is expressed as
\begin{equation}
Q=\sum^4_{j=1}q_jC_j,
\end{equation}
where $q_j$ are determined by
\begin{equation}
C_iQ=\lambda_iQ,
\end{equation}
that is,
\begin{equation}
\left[
D(C_i)-\lambda_i
\begin{pmatrix}
1 & 0 & 0 & 0\\
0 & 1 & 0 & 0\\
0 & 0 & 1 & 0\\
0 & 0 & 0 & 1
\end{pmatrix}
\right]
\begin{pmatrix}
q_1\\ q_2\\ q_3\\ q_4
\end{pmatrix}
=0.
\end{equation}
Solving the above equations, we get the projection operator which differs from
$Q$ by a constant:
\begin{equation}
P=q_4(C_1+C_2+C_3+C_4).
\end{equation}
Let $P$ operate on the following six color wave functions:
\begin{eqnarray}
\varphi_1 &=& r(1)r(2)\bar{r}(3)\bar{r}(4),\\
\varphi_2 &=& g(1)g(2)\bar{g}(3)\bar{g}(4),\\
\varphi_3 &=& b(1)b(2)\bar{b}(3)\bar{b}(4),\\
\varphi_4 &=& r(1)g(2)\bar{r}(3)\bar{g}(4),\\
\varphi_5 &=& g(1)b(2)\bar{g}(3)\bar{b}(4),\\
\varphi_6 &=& b(1)r(2)\bar{b}(3)\bar{r}(4),
\end{eqnarray}
and then add normalized $P\varphi_1$, $P\varphi_2$, $P\varphi_3$, $P\varphi_4$,
$P\varphi_5$, and $P\varphi_6$ to give the
color singlet of $q_1$, $\bar{q}_2$, $c$, and $\bar{c}$,
\begin{eqnarray}
\varphi_{q_1\bar{q}_2c\bar{c}{\rm color}}&=& 
\frac{1}{\sqrt{6}}
\left\{r(1)r(2)\bar{r}(3)\bar{r}(4)+g(1)g(2)\bar{g}(3)\bar{g}(4)
+b(1)b(2)\bar{b}(3)\bar{b}(4)\right.
	  \nonumber\\
&& +\frac{1}{2}
[ r(1)g(2)\bar{r}(3)\bar{g}(4)+r(2)g(1)\bar{r}(3)\bar{g}(4)
+r(1)g(2)\bar{r}(4)\bar{g}(3)
	  \nonumber\\
&& +r(2)g(1)\bar{r}(4)\bar{g}(3)]+\frac{1}{2}[ g(1)b(2)\bar{g}(3)\bar{b}(4)
+g(2)b(1)\bar{g}(3)\bar{b}(4)
	  \nonumber\\
&& +g(1)b(2)\bar{g}(4)\bar{b}(3)+g(2)b(1)\bar{g}(4)\bar{b}(3) ]
+\frac{1}{2}[ b(1)r(2)\bar{b}(3)\bar{r}(4)
	  \nonumber\\
&& \left. +b(2)r(1)\bar{b}(3)\bar{r}(4)+b(1)r(2)\bar{b}(4)\bar{r}(3)
+b(2)r(1)\bar{b}(4)\bar{r}(3) ]\right\} .
\end{eqnarray}
It satisfies
\begin{equation}
(\vec{\lambda}_{q_1}+\vec{\lambda}_{\bar{q}_2}+\vec{\lambda}_c
+\vec{\lambda}_{\bar{c}})\varphi_{q_1\bar{q}_2c\bar{c}{\rm color}}=0.
\end{equation}
Denote by $\phi_{q_1{\rm c}m}$ ($\phi_{\bar{q}_2{\rm c}m}$, $\phi_{c{\rm c}m}$,
$\phi_{\bar{c}{\rm c}m}$) the color wave function of $q_1$ ($\bar{q}_2$, $c$,
$\bar c$). When $m$ runs from 1 to 3, $\phi_{q_1{\rm c}m}$ and 
$\phi_{c{\rm c}m}$ are $r$, $g$, and $b$, and $\phi_{\bar{q}_2{\rm c}m}$ and
$\phi_{\bar{c}{\rm c}m}$ are $\bar r$, $\bar g$, and $\bar b$, respectively.
With this notation the color singlet is given by the short expression,
\begin{equation}
\varphi_{q_1\bar{q}_2c\bar{c}{\rm color}}=\frac{1}{2\sqrt 6}
\sum_{m=1}^3\sum_{n=1}^3 
(\phi_{q_1{\rm c}m}\phi_{c{\rm c}n}
\phi_{\bar{q}_2{\rm c}m}\phi_{\bar{c}{\rm c}n} 
+\phi_{q_1{\rm c}m}\phi_{c{\rm c}n}
\phi_{\bar{q}_2{\rm c}n}\phi_{\bar{c}{\rm c}m}).
\end{equation}

\newpage
\begin{figure}[htbp]
  \centering
  \includegraphics[scale=0.6]{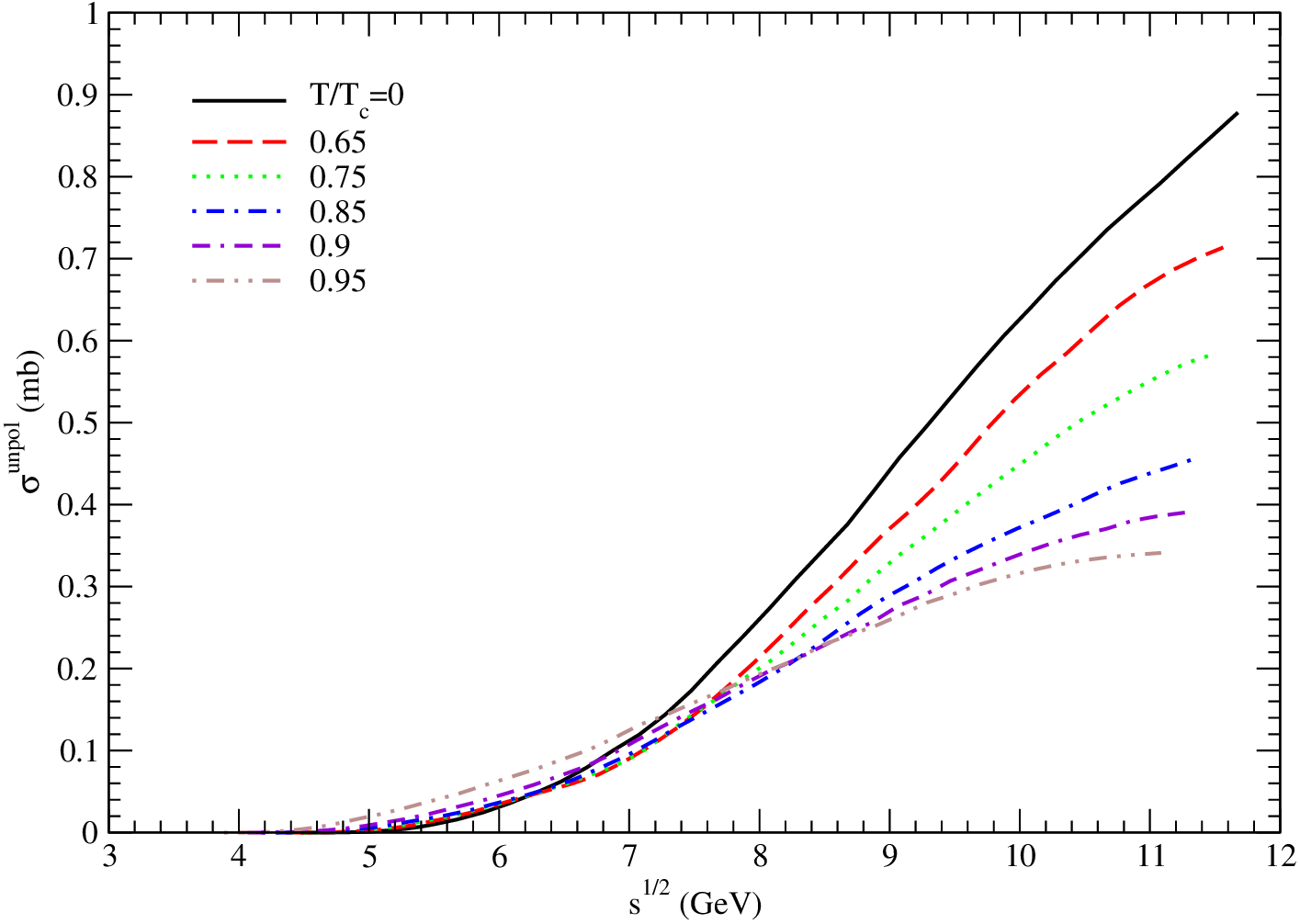}
  \caption{Cross sections for $\rho J/\psi \to D^+ X + D^0 X + D_s^+ X + 
D^{*+} X$ at various temperatures.}
\label{fig1}
\end{figure}

\newpage
\begin{figure}[htbp]
  \centering
  \includegraphics[scale=0.6]{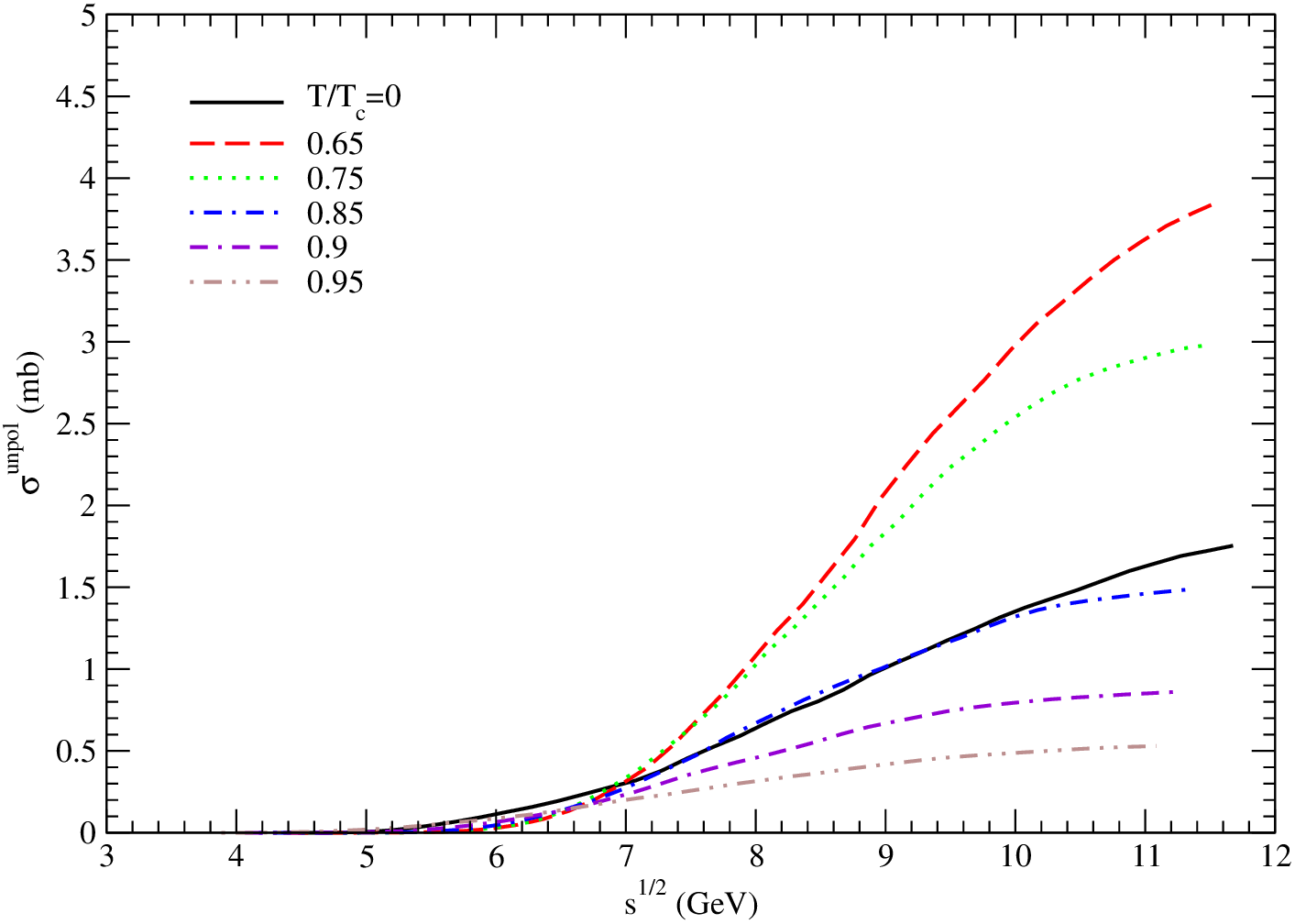}
  \caption{Cross sections for $\rho \psi' \to D^+ X + D^0 X + D_s^+ X + 
D^{*+} X$ at various temperatures.}
\label{fig2}
\end{figure}

\newpage
\begin{figure}[htbp]
  \centering
  \includegraphics[scale=0.6]{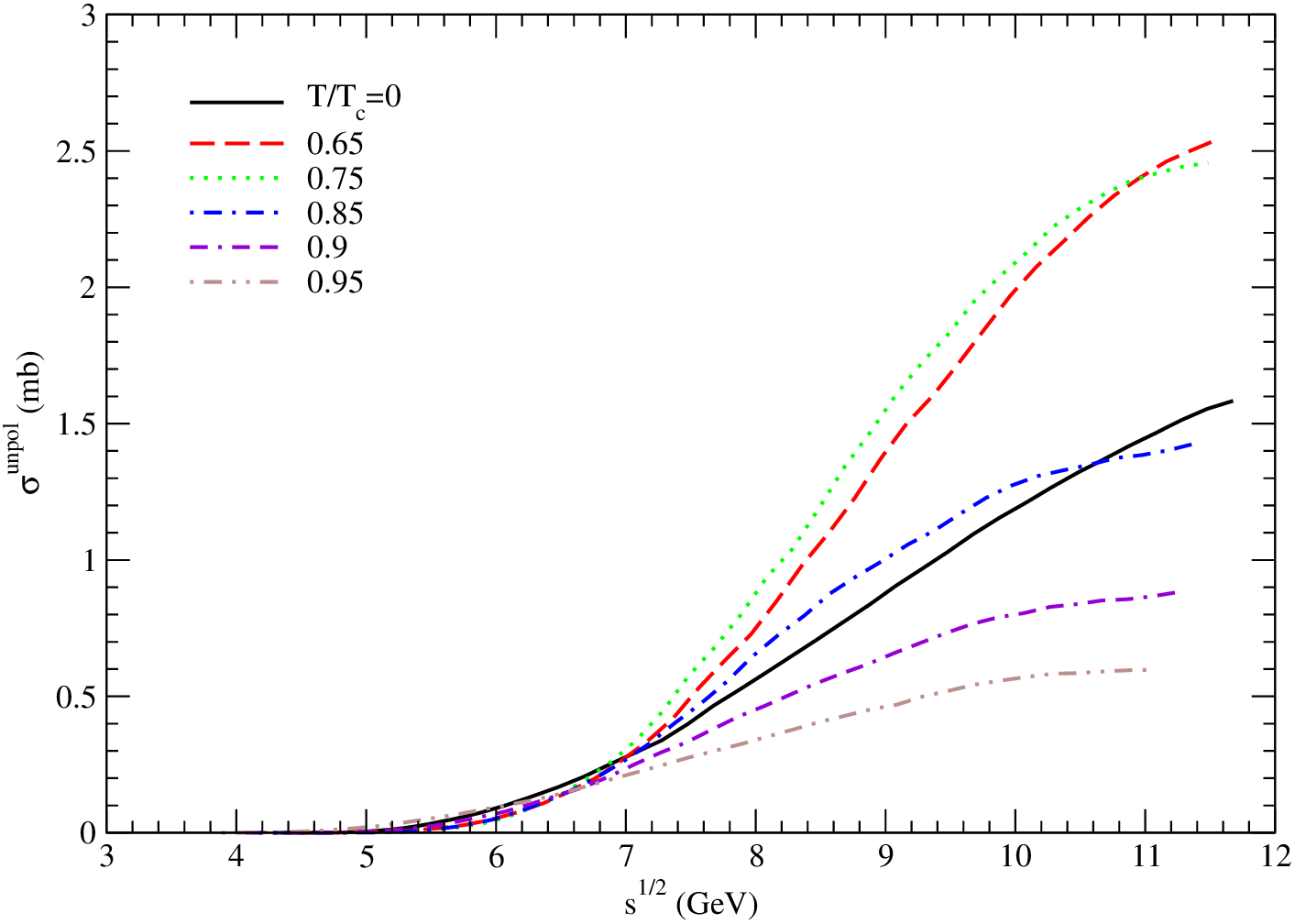}
  \caption{Cross sections for $\rho \chi_c \to D^+ X + D^0 X + D_s^+ X + 
D^{*+} X$ at various temperatures.}
\label{fig3}
\end{figure}


\newpage
\begin{figure}[htbp]
  \centering
  \includegraphics[scale=0.6]{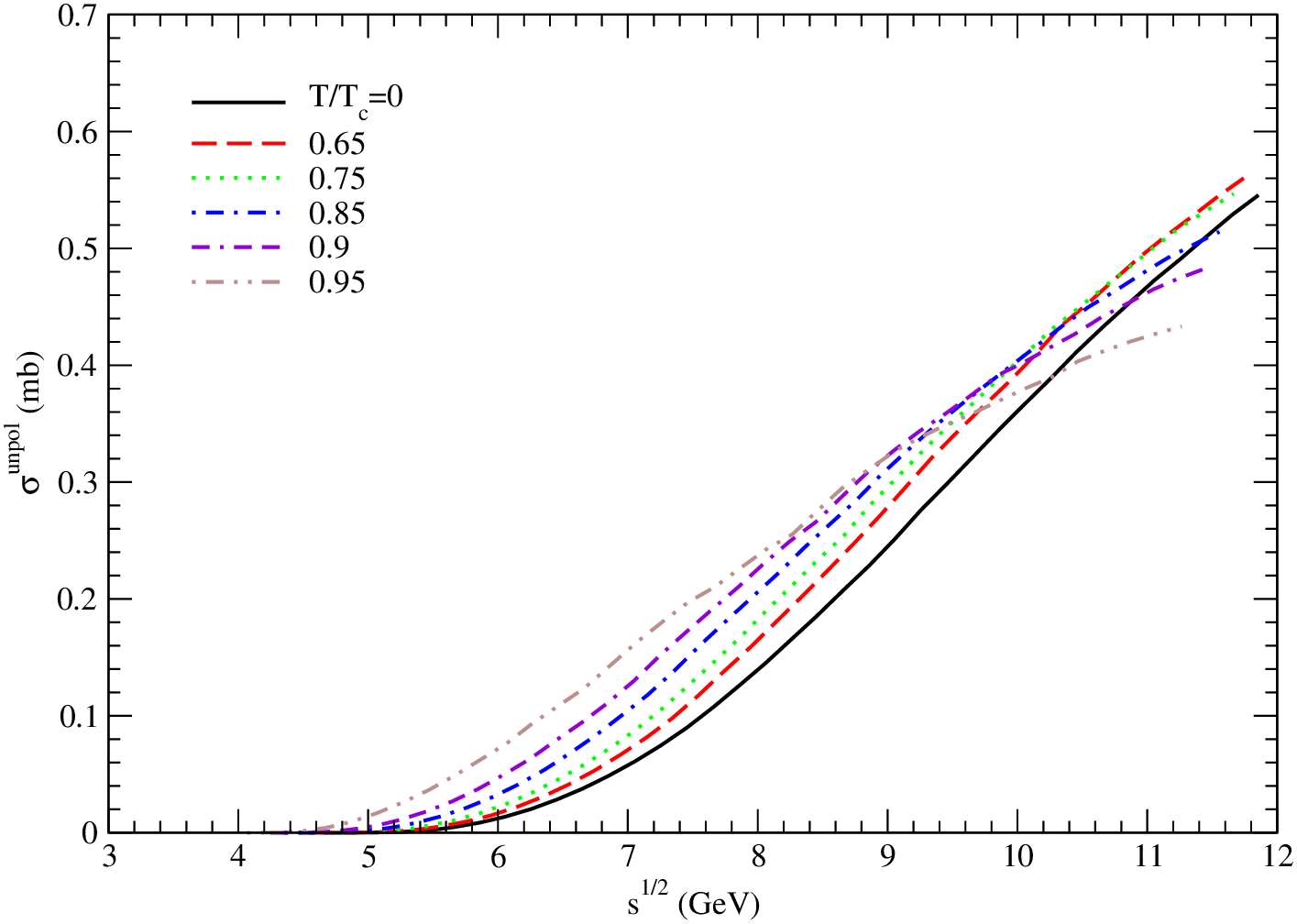}
  \caption{Cross sections for $K J/\psi \to D^+ X + D^0 X + D_s^+ X + D^{*+} X$
at various temperatures.}
\label{fig4}
\end{figure}

\newpage
\begin{figure}[htbp]
  \centering
  \includegraphics[scale=0.6]{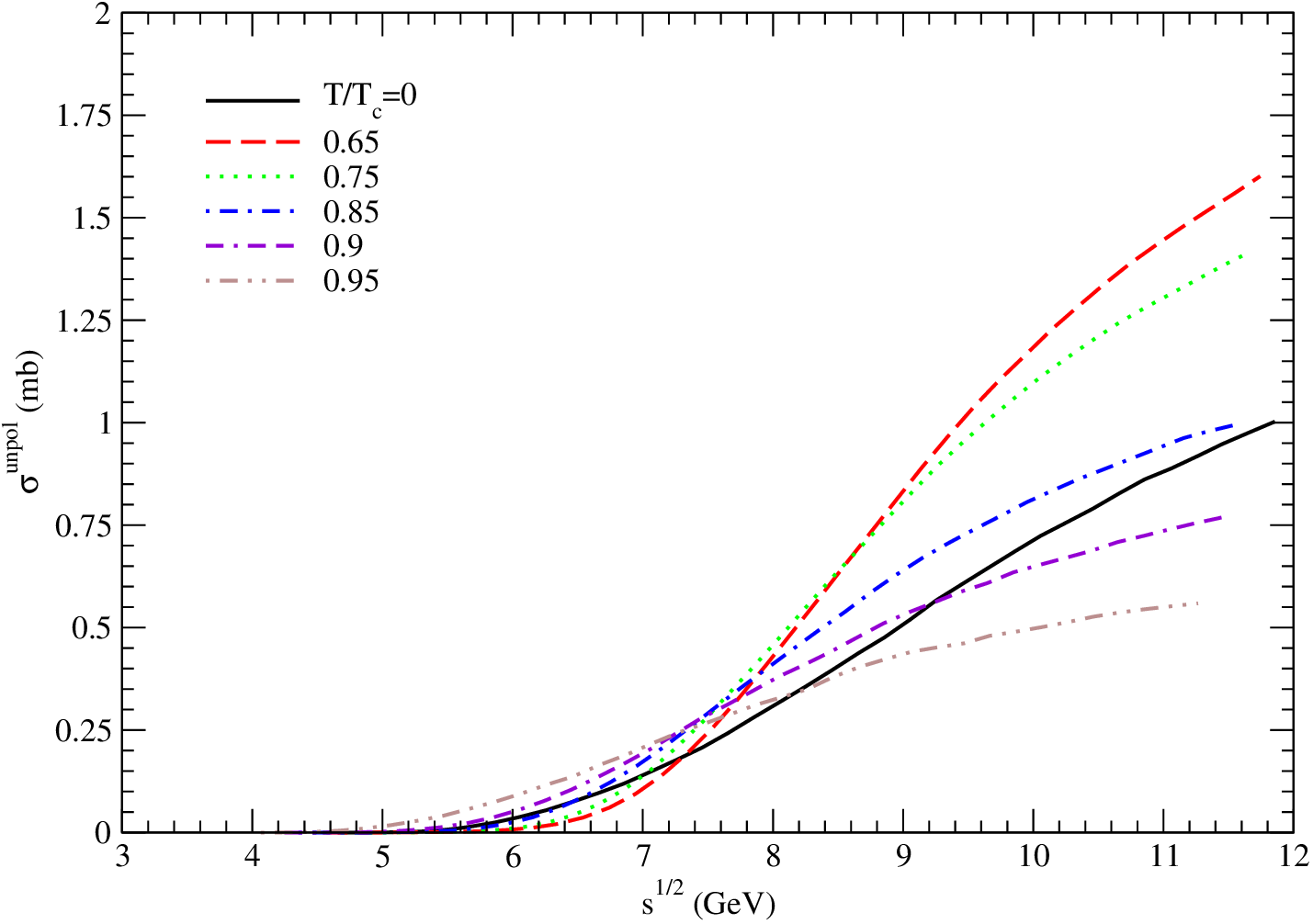}
  \caption{Cross sections for $K \psi' \to D^+ X + D^0 X + D_s^+ X + D^{*+} X$
at various temperatures.}
\label{fig5}
\end{figure}

\newpage
\begin{figure}[htbp]
  \centering
  \includegraphics[scale=0.6]{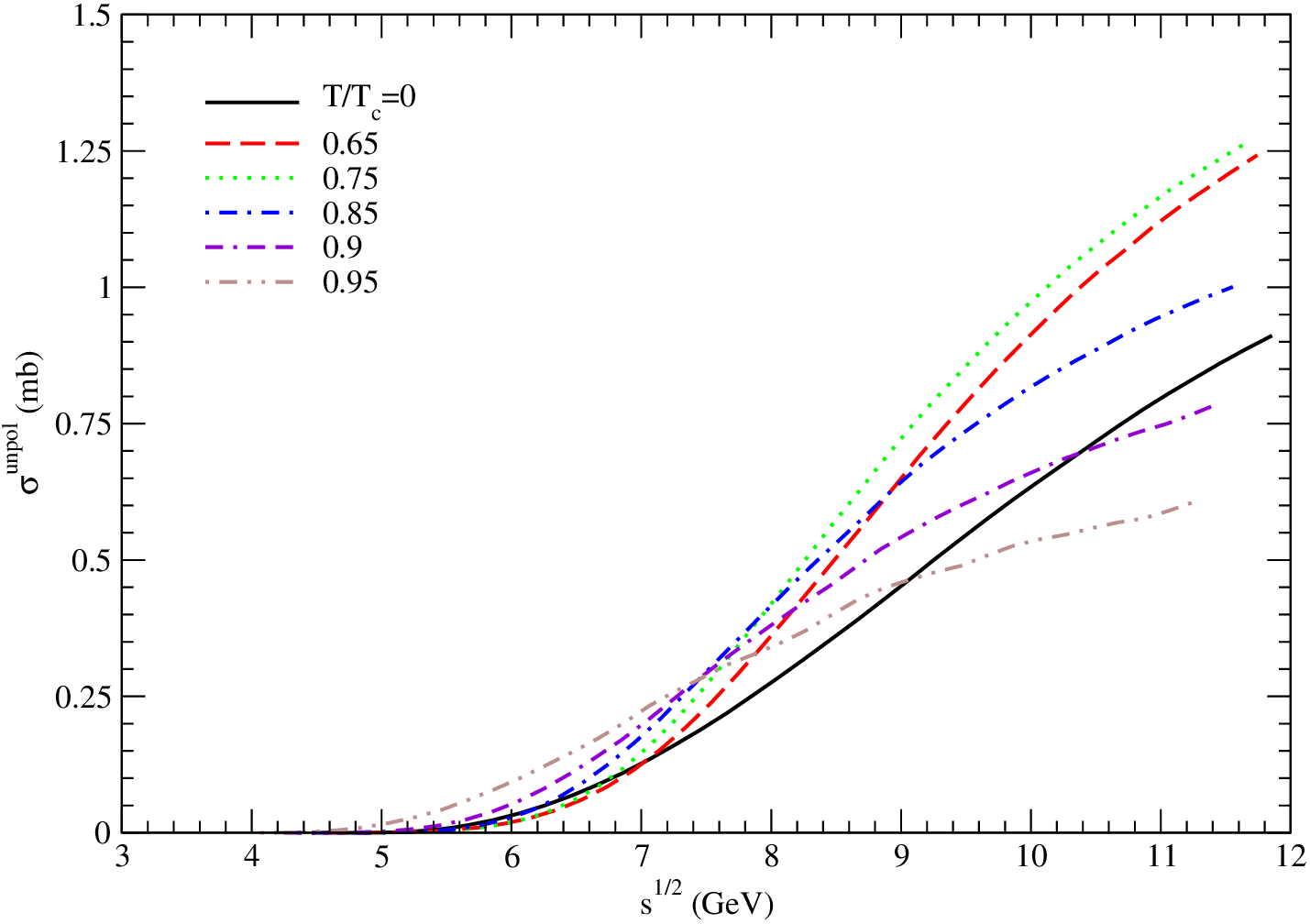}
  \caption{Cross sections for $K \chi_c \to D^+ X + D^0 X + D_s^+ X + D^{*+} X$
at various temperatures.}
\label{fig6}
\end{figure}


\newpage
\begin{figure}[htbp]
  \centering
  \includegraphics[scale=0.6]{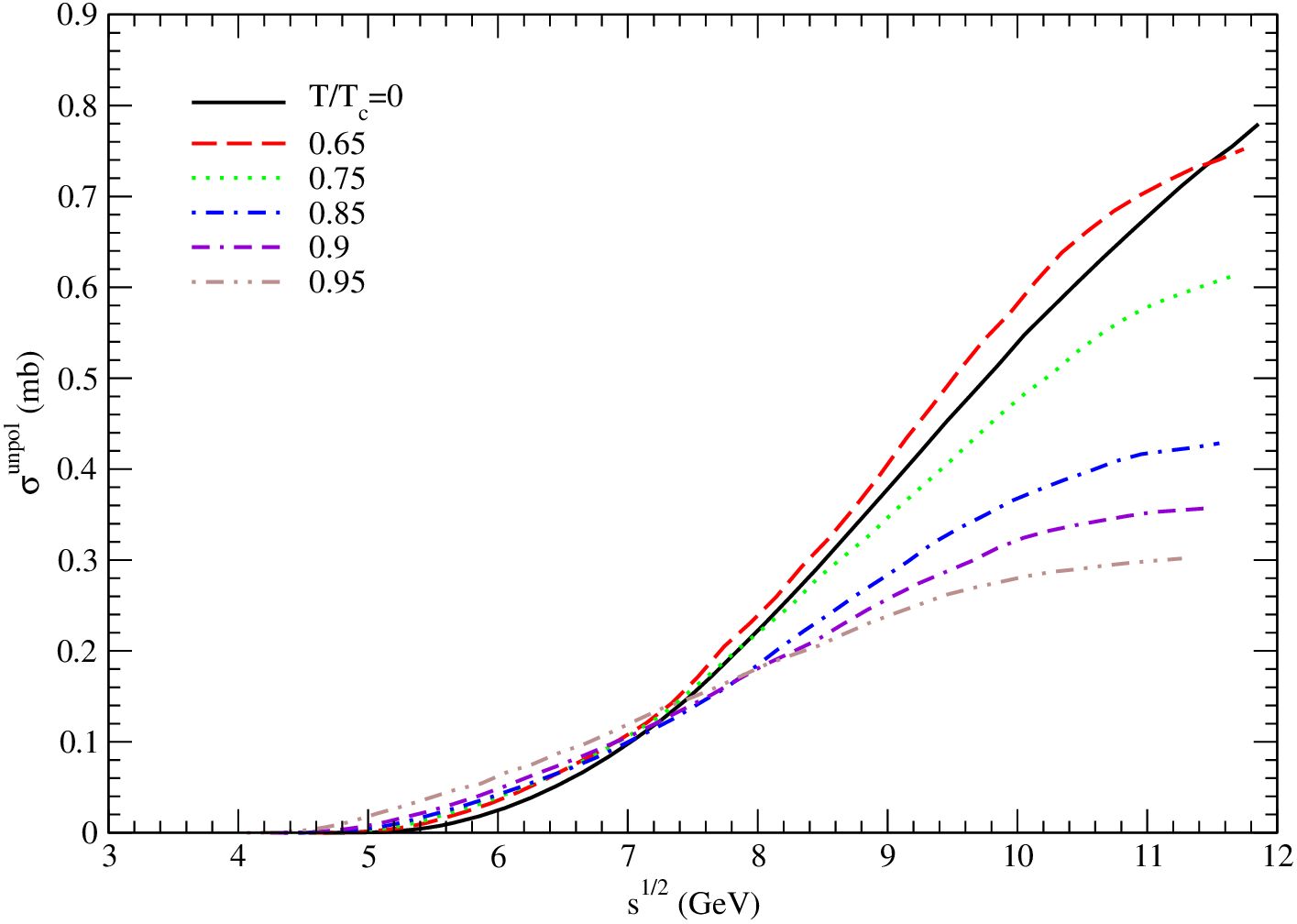}
  \caption{Cross sections for $K^\ast J/\psi \to D^+ X + D^0 X + D_s^+ X + 
D^{*+} X$ at various temperatures.}
\label{fig7}
\end{figure}

\newpage
\begin{figure}[htbp]
  \centering
  \includegraphics[scale=0.6]{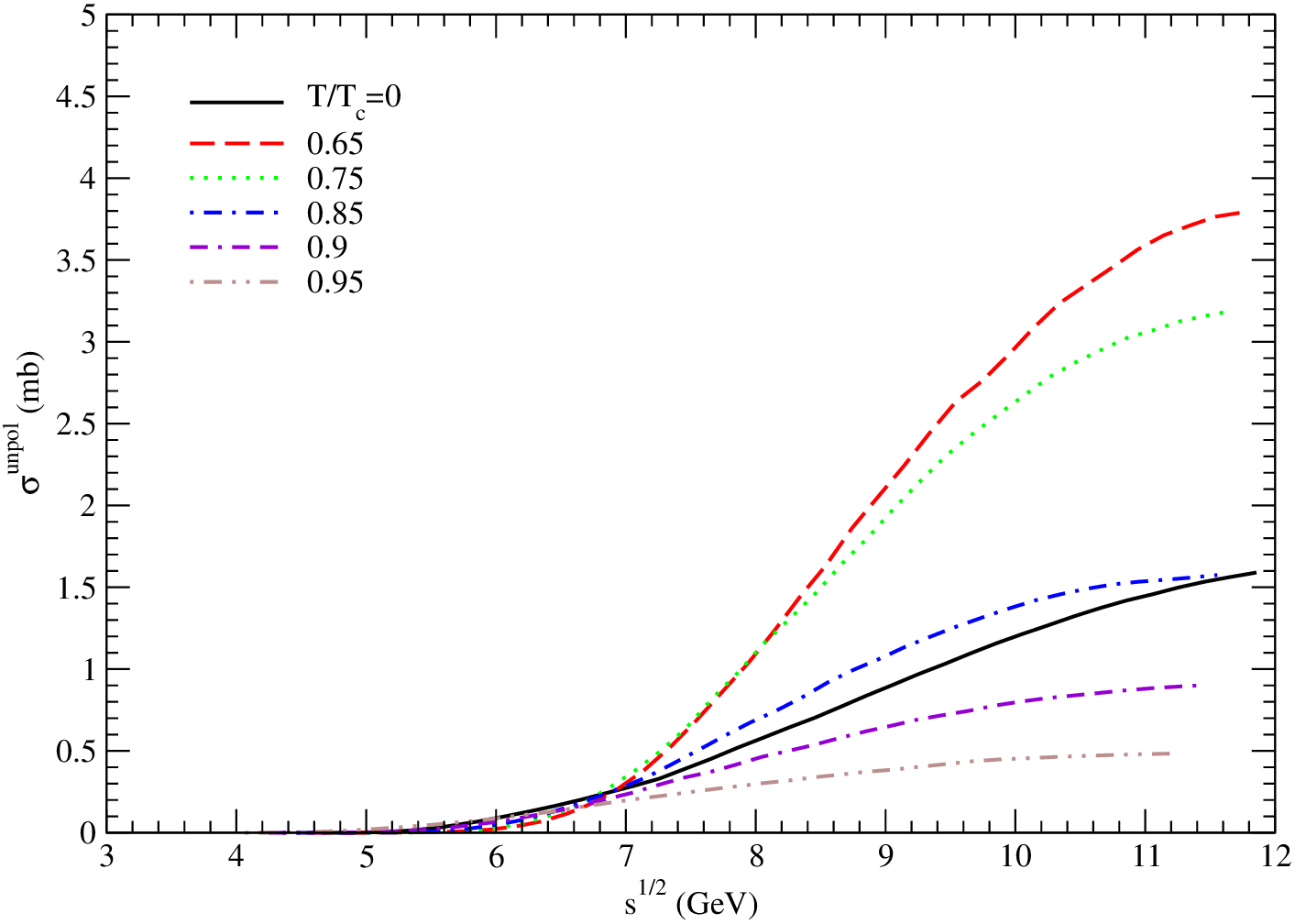}
  \caption{Cross sections for $K^\ast \psi' \to D^+ X + D^0 X + D_s^+ X + 
D^{*+} X$ at various temperatures.}
\label{fig8}
\end{figure}

\newpage
\begin{figure}[htbp]
  \centering
  \includegraphics[scale=0.6]{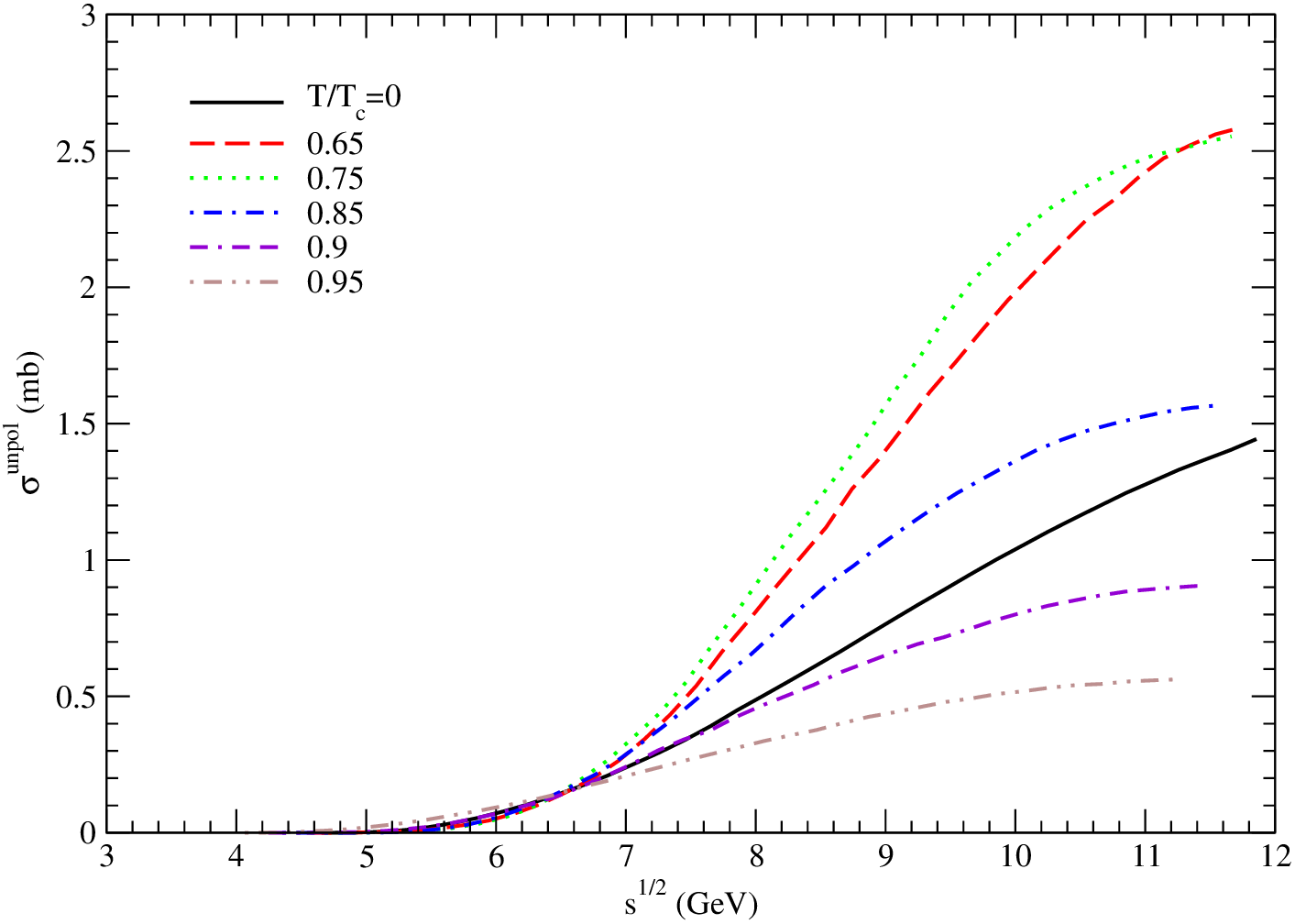}
  \caption{Cross sections for $K^\ast \chi_c \to D^+ X + D^0 X + D_s^+ X + 
D^{*+} X$ at various temperatures.}
\label{fig9}
\end{figure}


\newpage
\begin{figure}[htbp]
  \centering
  \includegraphics[scale=0.6]{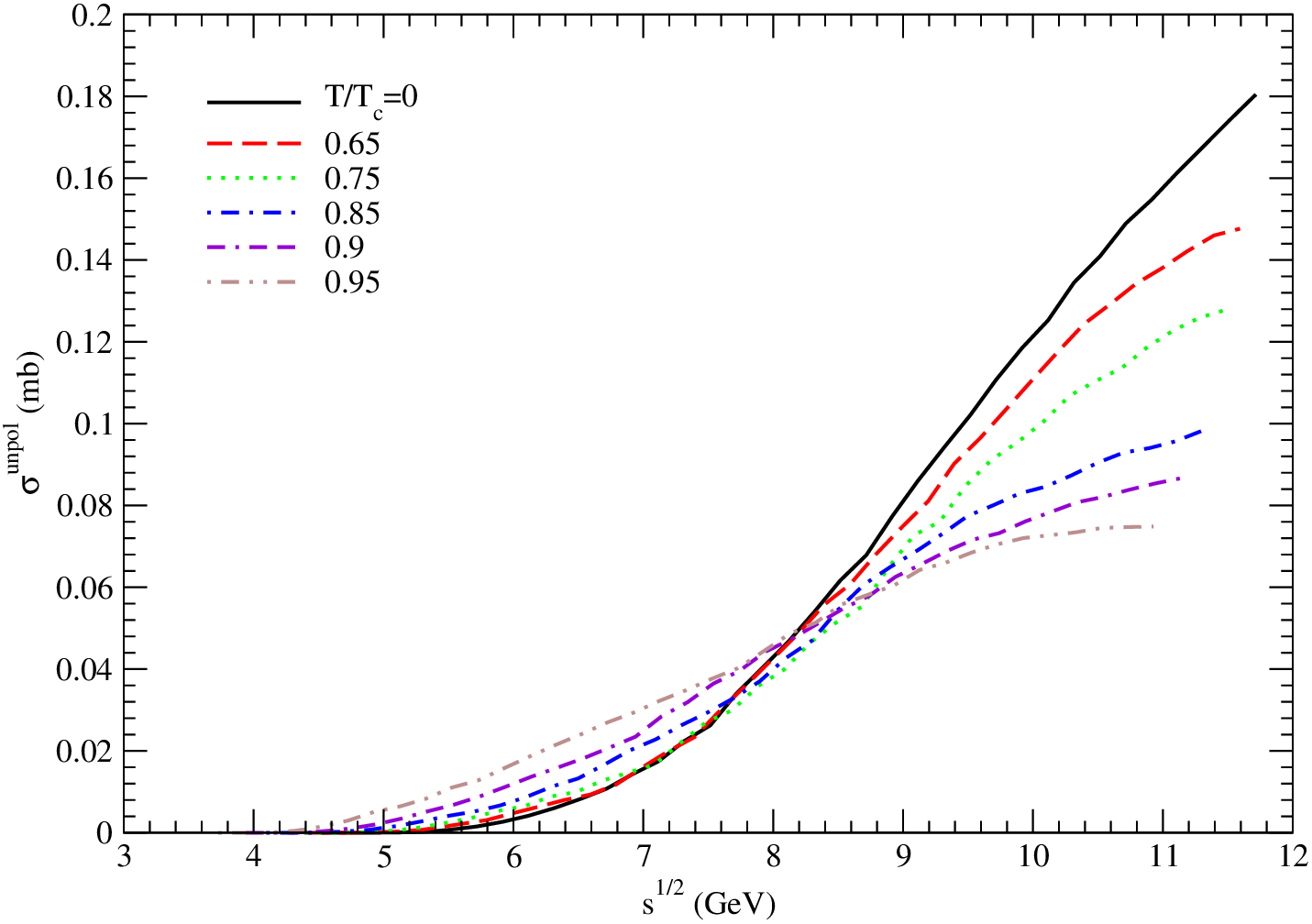}
  \caption{Cross sections for $\rho J/\psi\to D^{*+} X$ at various 
temperatures.}
\label{fig10}
\end{figure}

\newpage
\begin{figure}[htbp]
  \centering
  \includegraphics[scale=0.6]{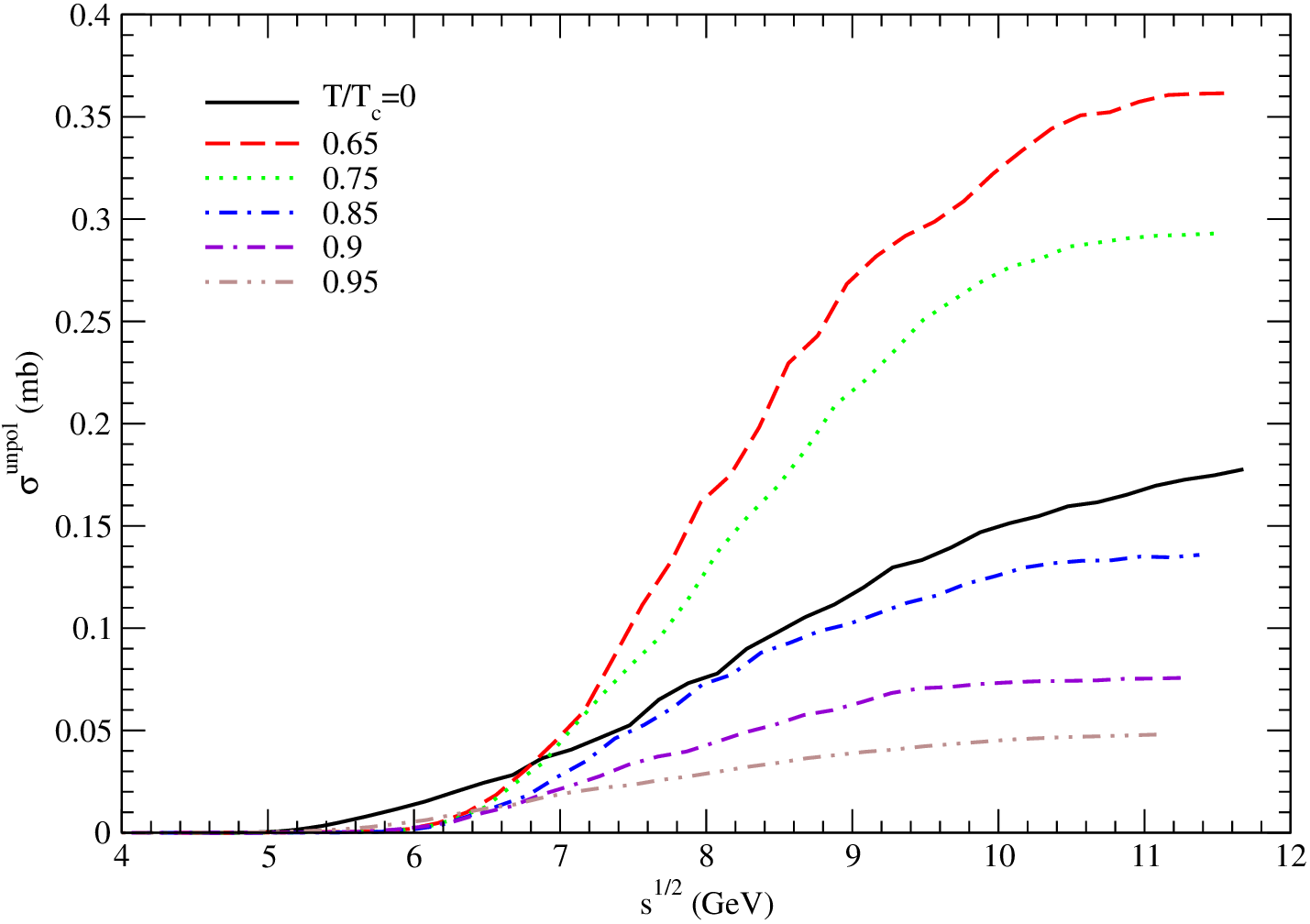}
  \caption{Cross sections for $\rho \psi'\to D^{+}_{s} X$ at various 
temperatures.}
\label{fig11}
\end{figure}

\newpage
\begin{figure}[htbp]
  \centering
  \includegraphics[scale=0.6]{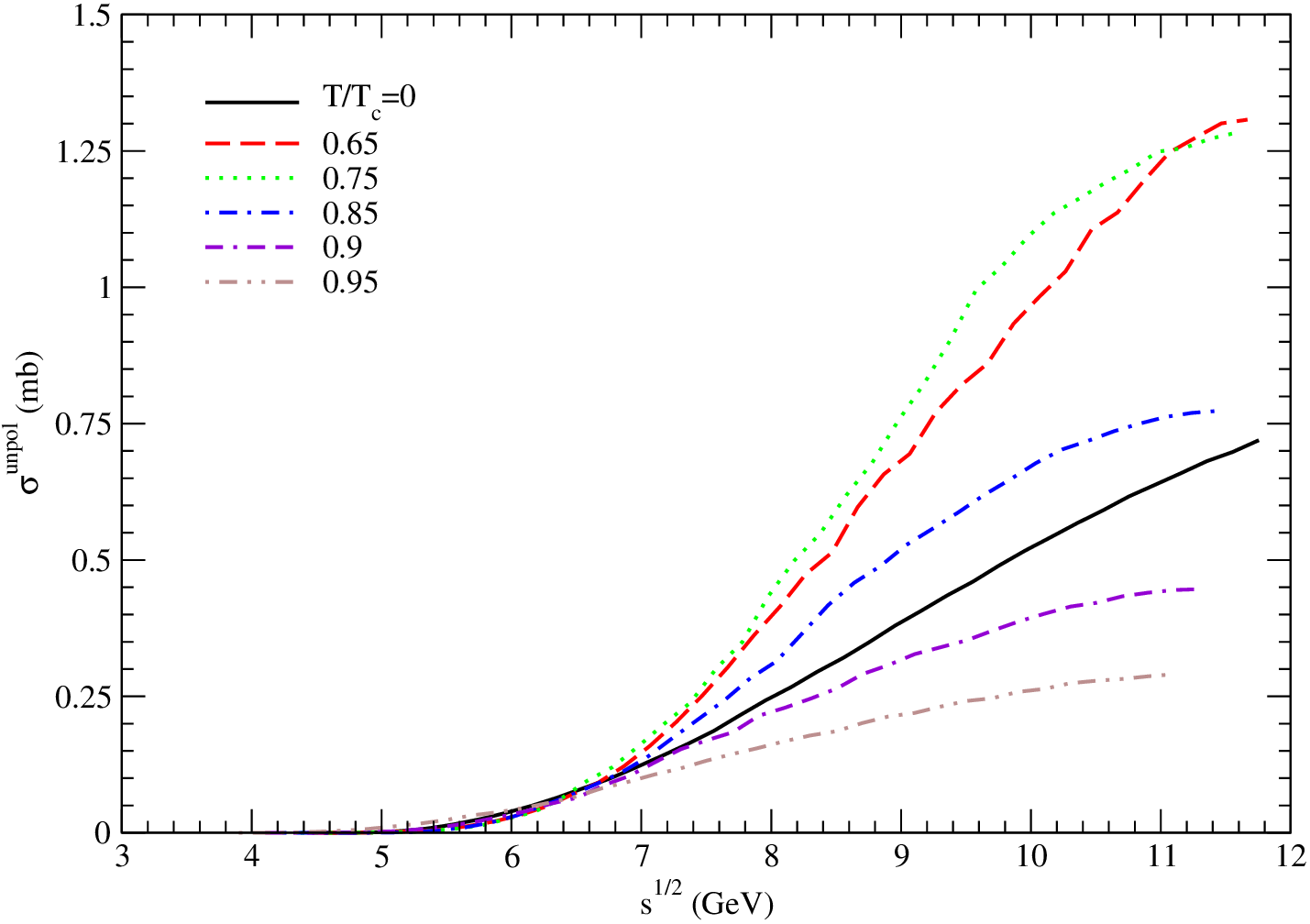}
  \caption{Cross sections for $K^{*} {\chi}_{c}\to D^{0} X$ at various 
temperatures.}
\label{fig12}
\end{figure}

\newpage

\begin{table}
\centering \caption{\label{table1} Spin matrix elements} 
\vspace{0.2cm}
\tabcolsep=5pt
\begin{tabular}{llllllllll}
  \hline
   $S_A$ & 0  & 0  & 0  & 1  & 1  & 1  & 1  & 1 & 1\\
   $S_B$ & 1  & 1  & 1  & 1  & 1  & 1  & 1  & 1 & 1\\
   $S^{\prime}_{q_1+\bar{q}_2}$ & 0  & 1  & 1  & 0  & 0  & 1  & 1  & 1 & 1\\
   $S^{\prime}_{c+\bar{c}}$  & 1  & 0  & 1  & 0  & 1  & 0  & 1  & 1 & 1\\
   $S$   & 1  & 1  & 1  & 0  & 1  & 1  & 0  & 1 & 2\\
  \hline
   $\psi^{SS_z+}_{\rm final}(\chi_A\chi_B)^S_{S_z}$ & 1  & 0  & 0  & 0  & 0  & 
0  & 1  & 1 & 1\\
   $\psi^{SS_z+}_{\rm final}\vec{s}_{q_1}\cdot\vec{s}_{\bar{c}}
(\chi_A\chi_B)^S_{S_z}$ 
   & 0  & $-\frac{1}{4}$ & $-\frac{1}{2\sqrt{2}}$  & $\frac{\sqrt{3}}{4}$ & 
$-\frac{1}{2\sqrt{2}}$  & $-\frac{1}{2\sqrt{2}}$  & $-\frac{1}{2}$  & 
$-\frac{1}{4}$ & $\frac{1}{4}$\\
   $\psi^{SS_z+}_{\rm final}\vec{s}_{\bar{q}_2}\cdot\vec{s}_{c}
(\chi_A\chi_B)^S_{S_z}$ 
   & 0  & $-\frac{1}{4}$ & $\frac{1}{2\sqrt{2}}$  & $\frac{\sqrt{3}}{4}$ & 
$\frac{1}{2\sqrt{2}}$  & $\frac{1}{2\sqrt{2}}$  & $-\frac{1}{2}$  & 
$-\frac{1}{4}$ & $\frac{1}{4}$\\
   $\psi^{SS_z+}_{\rm final}\vec{s}_{q_1}\cdot\vec{s}_{c}
(\chi_A\chi_B)^S_{S_z}$ 
   & 0  & $\frac{1}{4}$ & $-\frac{1}{2\sqrt{2}}$  & $-\frac{\sqrt{3}}{4}$ & 
$-\frac{1}{2\sqrt{2}}$  & $\frac{1}{2\sqrt{2}}$  & $-\frac{1}{2}$  & 
$-\frac{1}{4}$ & $\frac{1}{4}$\\
   $\psi^{SS_z+}_{\rm final}\vec{s}_{\bar{q}_2}\cdot\vec{s}_{\bar{c}}
(\chi_A\chi_B)^S_{S_z}$ 
   & 0  & $\frac{1}{4}$ & $\frac{1}{2\sqrt{2}}$  & $-\frac{\sqrt{3}}{4}$ & 
$\frac{1}{2\sqrt{2}}$  & $-\frac{1}{2\sqrt{2}}$  & $-\frac{1}{2}$  & 
$-\frac{1}{4}$ & $\frac{1}{4}$\\
  \hline
\end{tabular}
\end{table}

\begin{table}
\centering 
\caption{\label{table2} Quantities relevant to the cross sections for
the $\rho J/\psi$ dissociation reactions. $a_1$ and $a_2$ are in
units of mb; $b_1$, $b_2$, $d_0$, and $\sqrt{s_{\rm z}}$ are in
units of GeV; $c_1$ and $c_2$ are dimensionless.} 
\vspace{0.2cm}
\tabcolsep=5pt
\begin{tabular}{llllllllll}
  \hline
  Reaction & $T/T_{\rm c}$ & $a_1$ & $b_1$ & $c_1$ & $a_2$ & $b_2$ & $c_2$ 
&$d_0$ & $\sqrt{s_{\rm z}}$\\
  \hline
$\rho+J/{\psi} \to D^{+}+X$
  & 0 & 0.02 & 4 & 4.1 & 0.16 & 8.3 & 6.2 & 8.11 & 27.19\\
  & 0.65 & 0.01 & 2.9 & 4.2 & 0.13 & 7.2 & 7.9 & 7.16 & 22.27\\
  & 0.75 & 0.006 & 2.5 & 4.4 & 0.106 & 7.7 & 5.7 & 7.69 & 26.2\\
  & 0.85 & 0.002 & 1.8 & 5.7 & 0.084 & 8.2 & 4.2 & 8.2 & 31.01\\
  & 0.9  & 0.002 & 2 & 5.1 & 0.082 & 8.7 & 3.8 & 8.7 & 33.94\\
  & 0.95 & 0.002 & 1.9 & 6.4 & 0.068 & 8.7 & 3.2 & 8.7 & 36.47\\
$\rho+J/{\psi} \to D^{0}+X$
  & 0 & 0.05 & 4 & 4.1 & 0.43 & 8.3 & 6.3 & 8.13 & 27.04\\
  & 0.65 & 0.02 & 2.4 & 5.3 & 0.34 & 7.6 & 6.8 & 7.6 & 24.45\\
  & 0.75 & 0.007 & 1.8 & 6.4 & 0.302 & 8.6 & 4.4 & 8.6 & 31.84\\
  & 0.85 & 0.003 & 1.5 & 9.7 & 0.23 & 8.8 & 3.7 & 8.8 & 34.83\\
  & 0.9  & 0.004 & 1.8 & 5.9 & 0.186 & 8 & 3.9 & 8 & 31.18\\
  & 0.95 & 0.008 & 1.9 & 5.9 & 0.162 & 7.6 & 3.9 & 7.6 & 29.61\\
$\rho+J/{\psi} \to D^{+}_{s}+X$
  & 0 & 0.007 & 3.1 & 4.6 & 0.084 & 7.3 & 5.7 & 7.23 & 25.32\\
  & 0.65 & 0.003 & 1.9 & 5.7 & 0.066 & 6.9 & 5.2 & 6.9 & 24.9\\
  & 0.75 & 0.01 & 3.4 & 4.2 & 0.051 & 6.7 & 8.9 & 6.51 & 20.21\\
  & 0.85 & 0.001 & 2 & 4.9 & 0.042 & 7.5 & 4.2 & 7.5 & 28.82\\
  & 0.9  & 0.001 & 2.9 & 4.5 & 0.038 & 7.8 & 4 & 7.78 & 30.31\\
  & 0.95 & 0.004 & 3.7 & 4.3 & 0.031 & 8 & 4.3 & 7.75 & 29.81\\
$\rho+J/{\psi} \to D^{*+}+X$
  & 0 & 0.008 & 3.1 & 5.6 & 0.181 & 8 & 6.3 & 7.99 & 26.42\\
  & 0.65 & 0.002 & 1.7 & 7.2 & 0.156 & 8.5 & 4.8 & 8.5 & 30.63\\
  & 0.75 & 0.005 & 2.2 & 5.8 & 0.123 & 7.6 & 6 & 7.6 & 25.53\\
  & 0.85 & 0.05 & 7.2 & 10.4 & 0.05 & 7.4 & 3 & 7.24 & 30.58\\
  & 0.9  & 0.003 & 1.9 & 5.8 & 0.086 & 7.9 & 3.8 & 7.9 & 31.2\\
  & 0.95 & 0.002 & 1.6 & 7.8 & 0.071 & 7.5 & 3.1 & 7.5 & 32.44\\
  \hline
\end{tabular}
\end{table}

\begin{table}
\centering \caption{\label{table3} The same as Table 2 except for $\rho\psi'$.}
\vspace{0.2cm}
\tabcolsep=5pt
\begin{tabular}{llllllllll}
  \hline
  Reaction & $T/T_{\rm c}$ & $a_1$ & $b_1$ & $c_1$ & $a_2$ & $b_2$ & $c_2$ 
&$d_0$ & $\sqrt{s_{\rm z}}$\\
  \hline
$\rho+{\psi}^{\prime} \to D^{+}+X$
  & 0 & 0.01 & 1.7 & 5.3 & 0.347 & 8.4 & 3.6 & 8.4 & 34.23\\
  & 0.65 & 0.1 & 5 & 9 & 0.7 & 8.2 & 6.4 & 7.88 & 26.51\\
  & 0.75 & 0.2 & 5 & 9 & 0.6 & 9.7 & 7.3 & 9.3 & 29.13\\
  & 0.85 & 0.02 & 3.2 & 12.5 & 0.28 & 7.3 & 6.8 & 7.3 & 23.42\\
  & 0.9  & 0.003 & 2.5 & 12.9 & 0.162 & 6.9 & 5.5 & 6.9 & 23.96\\
  & 0.95 & 0.002 & 2 & 4.9 & 0.098 & 7.3 & 4 & 7.3 & 28.28\\
$\rho+{\psi}^{\prime} \to D^{0}+X$
  & 0 & 0.1 & 2.9 & 3.8 & 0.816 & 7.1 & 6.6 & 7.01 & 23.5\\
  & 0.65 & 0.2 & 5 & 8 & 1.8 & 7.9 & 7 & 7.67 & 24.92\\
  & 0.75 & 0.11 & 3.9 & 19.2 & 1.71 & 8.3 & 6.2 & 8.3 & 27.08\\
  & 0.85 & 0.11 & 3.9 & 9.2 & 0.72 & 7 & 9.2 & 6.86 & 20.54\\
  & 0.9  & 0.02 & 3 & 8 & 0.42 & 6.8 & 6.3 & 6.79 & 22.54\\
  & 0.95 & 0.01 & 2.6 & 3.9 & 0.26 & 7.6 & 4.2 & 7.58 & 28.69\\
$\rho+{\psi}^{\prime} \to D^{+}_{s}+X$
  & 0 & 0.01 & 2 & 5.1 & 0.17 & 6.9 & 4.7 & 6.9 & 25.98\\
  & 0.65 & 0.12 & 4.2 & 12 & 0.332 & 7.1 & 10.3 & 6.72 & 20.37\\
  & 0.75 & 0.04 & 3.4 & 19.6 & 0.299 & 6.4 & 9.9 & 6.39 & 18.98\\
  & 0.85 & 0.022 & 3.3 & 20.6 & 0.139 & 6.2 & 10.4 & 6.19 & 18.16\\
  & 0.9  & 0.011 & 3.3 & 21.1 & 0.077 & 6.4 & 7.7 & 6.39 & 20.22\\
  & 0.95 & 0.005 & 3.2 & 12.8 & 0.048 & 6.9 & 5.3 & 6.89 & 24.25\\
$\rho+{\psi}^{\prime} \to D^{*+}+X$
  & 0 & 0.01 & 2 & 6.5 & 0.35 & 7.7 & 5 & 7.7 & 27.84\\
  & 0.65 & 0.031 & 3.5 & 29.8 & 0.79 & 6.9 & 9.7 & 6.9 & 20.36\\
  & 0.75 & 0.11 & 3.9 & 14.7 & 0.61 & 6.6 & 15 & 6.52 & 17.47\\
  & 0.85 & 0.07 & 4.2 & 10.5 & 0.33 & 7.9 & 6.7 & 7.7 & 25.15\\
  & 0.9  & 0.01 & 3 & 13 & 0.18 & 6.6 & 5.4 & 6.59 & 23.26\\
  & 0.95 & 0.01 & 3.5 & 3.9 & 0.11 & 7.8 & 3.5 & 7.6 & 31.72\\
  \hline
\end{tabular}
\end{table}

\begin{table}
\centering 
\caption{\label{table4} The same as Table 2 except for $\rho\chi_c$.}
\vspace{0.2cm}
\tabcolsep=5pt
\begin{tabular}{llllllllll}
  \hline
  Reaction & $T/T_{\rm c}$ & $a_1$ & $b_1$ & $c_1$ & $a_2$ & $b_2$ & $c_2$ 
&$d_0$ & $\sqrt{s_{\rm z}}$\\
  \hline
$\rho+{\chi}_{c} \to D^{+}+X$
  & 0 & 0.01 & 2.3 & 4.1 & 0.31 & 8.5 & 3.9 & 8.5 & 33.33\\
  & 0.65 & 0.014 & 2.5 & 9.6 & 0.451 & 7.1 & 7 & 7.1 & 22.9\\
  & 0.75 & 0.1 & 6 & 6 & 0.4 & 8.1 & 5.8 & 7.62 & 26.8\\
  & 0.85 & 0.02 & 4 & 9 & 0.28 & 7.7 & 5.6 & 7.62 & 26.22\\
  & 0.9  & 0.02 & 4 & 7 & 0.18 & 8.7 & 4.7 & 8.59 & 31.02\\
  & 0.95 & 0.002 & 2 & 7.4 & 0.121 & 8.3 & 3.6 & 8.3 & 33.14\\
$\rho+{\chi}_{c} \to D^{0}+X$
  & 0 & 0.1 & 4 & 3.4 & 0.96 & 10.2 & 4.3 & 10.03 & 37.47\\
  & 0.65 & 0.01 & 7 & 2 & 1.25 & 8.1 & 5.1 & 8.1 & 28.75\\
  & 0.75 & 0.1 & 3.8 & 8.4 & 1.3 & 7.5 & 7.5 & 7.44 & 23.31\\
  & 0.85 & 0.1 & 4.1 & 7.5 & 0.7 & 7.5 & 7.8 & 7.33 & 22.85\\
  & 0.9  & 0.01 & 3 & 10 & 0.4 & 7.2 & 5 & 7.2 & 25.74\\
  & 0.95 & 0.02 & 3 & 4 & 0.3 & 7.8 & 4.7 & 7.74 & 28.02\\
$\rho+{\chi}_{c} \to D^{+}_{s}+X$
  & 0 & 0.01 & 3 & 4.1 & 0.16 & 8 & 3.8 & 7.94 & 32.07\\
  & 0.65 & 0.005 & 5 & 2.9 & 0.232 & 6.5 & 6 & 6.48 & 22.57\\
  & 0.75 & 0.014 & 3.9 & 15 & 0.247 & 6.4 & 7.1 & 6.34 & 20.96\\
  & 0.85 & 0.02 & 3.4 & 16.1 & 0.14 & 6.4 & 8.8 & 6.37 & 19.52\\
  & 0.9  & 0.01 & 3.5 & 17.1 & 0.08 & 6.4 & 7.2 & 6.36 & 20.65\\
  & 0.95 & 0.013 & 3.7 & 8.8 & 0.049 & 7.1 & 7.1 & 6.85 & 22.35\\
$\rho+{\chi}_{c} \to D^{*+}+X$
  & 0 & 0.01 & 2.6 & 5.6 & 0.33 & 8.1 & 4.9 & 8.1 & 29.28\\
  & 0.65 & 0.1 & 5.5 & 11.4 & 0.6 & 10.1 & 4.9 & 9.8 & 35.23\\
  & 0.75 & 0.1 & 4.9 & 8.4 & 0.5 & 7.6 & 8.4 & 7.25 & 22.7\\
  & 0.85 & 0.012 & 3.1 & 13.1 & 0.305 & 6.3 & 7.3 & 6.29 & 20.34\\
  & 0.9  & 0.01 & 2.7 & 11.2 & 0.192 & 6.9 & 5.3 & 6.9 & 24.3\\
  & 0.95 & 0.03 & 3.4 & 4.4 & 0.12 & 7.1 & 7 & 6.81 & 22.25\\
  \hline
  \end{tabular}
\end{table}

\begin{table}
\centering \caption{\label{table5} The same as Table 2 except for $KJ/\psi$.}
\vspace{0.2cm}
\tabcolsep=5pt
\begin{tabular}{llllllllll}
  \hline
  Reaction & $T/T_{\rm c}$ & $a_1$ & $b_1$ & $c_1$ & $a_2$ & $b_2$ & $c_2$ 
&$d_0$ & $\sqrt{s_{\rm z}}$\\
  \hline
  $K+J/\psi\to D^++X$
  & 0     & 0.01  & 0.01  & 0.08  & 0.11  & 9.13  & 3.77  & 9.13  & 36.18\\
  & 0.65  & 0.01  & 0.01  & 0.07  & 0.11  & 8.83  & 3.8   & 8.83  & 34.93\\
  & 0.75  & 0.01 & 0.01 & 0.08 & 0.105 & 8.71 & 3.68 & 8.71 & 34.91\\
  & 0.85  & 0.01 & 0.01 & 0.09 & 0.097 & 8.42 & 3.57 & 8.42 & 34.21\\
  & 0.9   & 0.01 & 0.01 & 0.07 & 0.088 & 7.99 & 3.46 & 7.99 & 33\\
  & 0.95  & 0.01 & 2.9 & 4.4 & 0.08 & 7.7 & 4.7 & 7.62 & 27.88\\
  $K+J/\psi\to D^0+X$
  & 0     & 0.01  & 0.01  & 0.06  & 0.28  & 9     & 3.85  & 9     & 35.41\\
  & 0.65  & 0.01  & 0.01  & 0.05  & 0.28  & 8.73  & 3.87  & 8.73  & 34.32\\
  & 0.75  & 0.01 & 0.01 & 0.09 & 0.29 & 9.13 & 3.64 & 9.13 & 36.56\\
  & 0.85  & 0.01 & 0.01 & 0.08 & 0.26 & 8.62 & 3.58 & 8.62 & 34.88\\
  & 0.9   & 0.01 & 0.01 & 0.08 & 0.24 & 8.39 & 3.41 & 8.39 & 34.67\\
  & 0.95  & 0.02 & 2.9  & 4.4 & 0.22 & 8.3 & 4 & 8.25 & 31.82\\
  $K+J/\psi\to D_s^++X$
  & 0     & 0.01  & 0.01  & 0.08  & 0.05  & 7.38  & 4.05  & 7.38  & 29.34\\
  & 0.65  & 0.01  & 0.01  & 0.07  & 0.06  & 8.12  & 3.69  & 8.12  & 32.97\\
  & 0.75  & 0.01 & 0.01 & 0.09 & 0.05 & 7.25 & 3.84 & 7.25 & 29.35\\
  & 0.85  & 0.01 & 4.4 & 4.6 & 0.05 & 9.4 & 4.1 & 8.9 & 35.48\\
  & 0.9   & 0.01 & 4.1 & 4.7 & 0.04 & 8 & 5.3 & 7.48 & 27.69\\
  & 0.95  & 0.01 & 3.7 & 5.1 & 0.04 & 8.7 & 4.9 & 8.41 & 30.61\\
  $K+J/\psi\to D^{*+}+X$
  & 0     & 0.01  & 0.01  & 0.09  & 0.11  & 8.5   & 4.5   & 8.5   & 31.73\\
  & 0.65  & 0.01  & 0.01  & 0.07  & 0.12  & 8.43  & 4.32  & 8.43  & 31.91\\
  & 0.75  & 0.01 & 0.01 & 0.13 & 0.12 & 8.45 & 4.04 & 8.45 & 32.76\\
  & 0.85  & 0.01 & 0.01 & 0.1 & 0.107 & 7.76 & 3.71 & 7.76 & 31.37\\
  & 0.9   & 0.01 & 0.01 & 0.09 & 0.099 & 7.35 & 3.42 & 7.35 & 30.86\\
  & 0.95  & 0.01 & 2.5 & 4.7 & 0.09 & 7.1 & 4.2 & 7.06 & 27.23\\
  \hline
\end{tabular}
\end{table}

\begin{table}
\centering \caption{\label{table6} The same as Table 2 except for $K\psi'$.}
\vspace{0.2cm}
\tabcolsep=5pt
\begin{tabular}{llllllllll}
  \hline
  Reaction & $T/T_{\rm c}$ & $a_1$ & $b_1$ & $c_1$ & $a_2$ & $b_2$ & $c_2$ 
&$d_0$ & $\sqrt{s_{\rm z}}$\\
  \hline
  $K+\psi'\to D^++X$
  & 0     & 0.01  & 0.01  & 0.08  & 0.18  & 8.02  & 3.69  & 8.02  & 32.63\\
  & 0.65  & 0.01  & 0.01  & 0.03  & 0.28  & 6.78  & 7.22  & 6.78  & 22.02\\
  & 0.75  & 0.1 & 4.8 & 9 & 0.243 & 8.47 & 9.6 & 7.9 & 23.92\\
  & 0.85  & 0.02 & 3.7 & 10 & 0.188 & 7.8 & 5.8 & 7.76 & 26.36\\
  & 0.9   & 0.02 & 4 & 7 & 0.15 & 8.7 & 4.5 & 8.55 & 31.75\\
  & 0.95  & 0.02 & 4.2 & 3.8 & 0.1 & 8.7 & 4 & 8.15 & 32.99\\
  $K+\psi'\to D^0+X$
  & 0     & 0.01  & 0.01  & 0.06  & 0.48  & 8.19  & 3.69  & 8.19  & 33.23\\
  & 0.65  & 0.01  & 0.01  & 0.02  & 0.73  & 6.81  & 7.26  & 6.81  & 22.06\\
  & 0.75  & 0.14 & 4.3 & 10 & 0.705 & 8.13 & 7.67 & 7.96 & 24.9\\
  & 0.85  & 0.1 & 4.2 & 8 & 0.481 & 8.1 & 6.9 & 7.86 & 25.55\\
  & 0.9   & 0.1 & 4.7 & 6 & 0.44 & 10.6 & 4.4 & 10.25 & 38.11\\
  & 0.95  & 0.01 & 3 & 6 & 0.3 & 8.6 & 3.3 & 8.59 & 35.76\\
  $K+\psi'\to D_s^++X$
  & 0     & 0.01  & 0.01  & 0.08  & 0.1   & 7.48  & 3.63  & 7.48  & 31.06\\
  & 0.65  & 0.05  & 4.2   & 13.6  & 0.15  & 7.6   & 7.5   & 7.34  & 23.89\\
  & 0.75  & 0.03 & 3.7 & 19 & 0.133 & 6.83 & 7.67 & 6.79 & 21.72\\
  & 0.85  & 0.017 & 3.4 & 20 & 0.091 & 6.6 & 7.5 & 6.59 & 21.18\\
  & 0.9   & 0.014 & 3.5 & 15 & 0.067 & 6.9 & 6.4 & 6.86 & 23.01\\
  & 0.95  & 0.02 & 4.1 & 6.6 & 0.05 & 8.6 & 6.4 & 8.1 & 27.36\\
  $K+\psi'\to D^{*+}+X$
  & 0     & 0.01  & 2.6  & 6.4  & 0.2   & 7.7   & 5.7  & 7.7    & 26.7\\
  & 0.65  & 0.1   & 5    & 14   & 0.47  & 10.4  & 6.5  & 10.36  & 32.68\\
  & 0.75  & 0.05 & 4.1 & 18 & 0.318 & 7.94 & 7.41 & 7.92 & 24.81\\
  & 0.85  & 0.047 & 3.9 & 10 & 0.211 & 7.6 & 7.1 & 7.44 & 24.06\\
  & 0.9   & 0.04 & 4 & 7 & 0.18 & 9.2 & 4.4 & 9.01 & 33.66\\
  & 0.95  & 0.03 & 4 & 4 & 0.12 & 9.3 & 3.5 & 8.65 & 37.19\\
  \hline
\end{tabular}
\end{table}

\begin{table}
\centering \caption{\label{table7} The same as Table 2 except for $K\chi_c$.}
\vspace{0.2cm}
\tabcolsep=5pt
\begin{tabular}{llllllllll}
  \hline
  Reaction & $T/T_{\rm c}$ & $a_1$ & $b_1$ & $c_1$ & $a_2$ & $b_2$ & $c_2$ 
&$d_0$ & $\sqrt{s_{\rm z}}$\\
  \hline
  $K+\chi_c\to D^++X$
  & 0     & 0.01  & 0.01  & 0.06  & 0.18  & 8.93  & 3.37  & 8.93  & 37.29\\
  & 0.65  & 0.01  & 0.01  & 0.05  & 0.23  & 7.6   & 4.98  & 7.6   & 27.53\\
  & 0.75  & 0.1 & 5.5 & 6.4 & 0.273 & 11.4 & 5.8 & 10.65 & 36.57\\
  & 0.85  & 0.03 & 4 & 8 & 0.183 & 7.9 & 6.2 & 7.72 & 25.98\\
  & 0.9   & 0.02 & 4 & 6.7 & 0.15 & 8.6 & 4.5 & 8.43 & 31.43\\
  & 0.95  & 0.01 & 3.6 & 5 & 0.12 & 9 & 3.3 & 8.9 & 37.23\\
  $K+\chi_c\to D^0+X$
  & 0     & 0.01  & 0.01  & 0.05  & 0.46  & 8.85  & 3.42  & 8.85  & 36.75\\
  & 0.65  & 0.01  & 0.01  & 0.04  & 0.59  & 7.59  & 5.04  & 7.59  & 27.38\\
  & 0.75  & 0.1 & 4.7 & 8.1 & 0.695 & 9.51 & 4.97 & 9.31 & 33.23\\
  & 0.85  & 0.1 & 4.4 & 8 & 0.519 & 9.1 & 5.3 & 8.89 & 31.06\\
  & 0.9   & 0.1 & 4.3 & 5.9 & 0.37 & 8.4 & 6.5 & 7.96 & 26.69\\
  & 0.95  & 0.1 & 5.1 & 4 & 0.4 & 14 & 3.2 & 13.51 & 56.52\\
  $K+\chi_c\to D_s^++X$
  & 0     & 0.01  & 0.01  & 0.08  & 0.09  & 7.6   & 3.52  & 7.6   & 31.9\\
  & 0.65  & 0.01  & 0.01  & 0.06  & 0.12  & 6.59  & 5.26  & 6.59  & 24.06\\
  & 0.75  & 0.013 & 3.6 & 18 & 0.12 & 6.9 & 5.8 & 6.88 & 24.04\\
  & 0.85  & 0.032 & 3.9 & 12 & 0.089 & 7.4 & 8.2 & 7.2 & 22.55\\
  & 0.9   & 0.02 & 3.7 & 11.3 & 0.07 & 7.3 & 7.5 & 7.17 & 22.85\\
  & 0.95  & 0.01 & 3.2 & 9.5 & 0.05 & 6.7 & 6.2 & 6.61 & 22.52\\
  $K+\chi_c\to D^{*+}+X$
  & 0     & 0.01  & 0.01  & 0.08  & 0.18  & 8.29  & 4.11  & 8.29  & 32.22\\
  & 0.65  & 0.01  & 0.01  & 0.06  & 0.28  & 8.15  & 5.44  & 8.15  & 28.35\\
  & 0.75  & 0.1 & 5.3 & 8.4 & 0.303 & 10.92 & 5.49 & 10.46 & 36.12\\
  & 0.85  & 0.05 & 4 & 9 & 0.207 & 7.6 & 7.1 & 7.35 & 24.04\\
  & 0.9   & 0.03 & 3.8 & 7.2 & 0.17 & 8.2 & 4.5 & 8.01 & 30.17\\
  & 0.95  & 0.02 & 3.7 & 4.6 & 0.13 & 8.7 & 3.2 & 8.39 & 36.6\\
  \hline
  \end{tabular}
\end{table}

\begin{table}
\centering 
\caption{\label{table8} The same as Table 2 except for $K^{*}J/\psi$.}
\vspace{0.2cm}
\tabcolsep=5pt
\begin{tabular}{llllllllll}
  \hline
  Reaction & $T/T_{\rm c}$ & $a_1$ & $b_1$ & $c_1$ & $a_2$ & $b_2$ & $c_2$ 
&$d_0$ & $\sqrt{s_{\rm z}}$\\
  \hline
$K^{*}+J/{\psi} \to D^{+}+X$
  & 0 & 0.01 & 3 & 4.4 & 0.15 & 8.3 & 5 & 8.27 & 29.69\\
  & 0.65 & 0.009 & 2.5 & 4.3 & 0.154 & 8.1 & 5.4 & 8.09 & 28.17\\
  & 0.75 & 0.01 & 2.7 & 4 & 0.12 & 7.9 & 5.6 & 7.87 & 27.13\\
  & 0.85 & 0.002 & 1.6 & 6.6 & 0.083 & 8.6 & 3.4 & 8.6 & 35.62\\
  & 0.9  & 0.01 & 2.8 & 4.1 & 0.07 & 7.1 & 6.5 & 7.01 & 23.27\\
  & 0.95 & 0.003 & 2 & 6 & 0.062 & 8.6 & 2.9 & 8.6 & 38.01\\
$K^{*}+J/{\psi} \to D^{0}+X$
  & 0 & 0.1 & 6  & 3.4 & 0.41 & 11.3 & 4.3 & 10.43 & 40.86\\
  & 0.65 & 0.01 & 1.8 & 5.6 & 0.39 & 8.3 & 4.3 & 8.3 & 31.44\\
  & 0.75 & 0.01 & 1.9 & 4.8 & 0.31 & 8.2 & 4.2 & 8.2 & 31.32\\
  & 0.85 & 0.01 & 2 & 5 & 0.22 & 8.1 & 4.1 & 8.1 & 31.17\\
  & 0.9  & 0.03 & 3.1 & 3.9 & 0.17 & 7.2 & 7.1 & 7.05 & 22.83\\
  & 0.95 & 0.004 & 1.5 & 8 & 0.14 & 7.4 & 3.1 & 7.4 & 32.23\\
$K^{*}+J/{\psi} \to D^{+}_{s}+X$
  & 0 & 0.02 & 4.3 & 4.1 & 0.07 & 8.2 & 6.5 & 7.7 & 26.61\\
  & 0.65 & 0.005 & 2.3 & 4.2 & 0.073 & 6.9 & 4.9 & 6.88 & 25.64\\
  & 0.75 & 0.007 & 3.2 & 3.5 & 0.057 & 7.8 & 4.3 & 7.62 & 29.76\\
  & 0.85 & 0.001 & 2 & 5.5 & 0.046 & 8 & 3.4 & 8 & 33.55\\
  & 0.9  & 0.004 & 2.6 & 5.1 & 0.031 & 6.9 & 5.3 & 6.85 & 24.6\\
  & 0.95 & 0.006 & 3 & 4.8 & 0.024 & 6.6 & 6.8 & 6.38 & 21.52\\
$K^{*}+J/{\psi} \to D^{*+}+X$
  & 0  & 0.03 & 4.7 & 4.6 & 0.15 & 8.4 & 7.5 & 8.07 & 25.97\\
  & 0.65 & 0.007 & 2.5 & 4.9 & 0.147 & 7.4 & 5.8 & 7.39 & 25.56\\
  & 0.75 & 0.01 & 2.7 & 4.7 & 0.12 & 7.2 & 6.2 & 7.17 & 24.31\\
  & 0.85 & 0.003 & 1.7 & 6.6 & 0.094 & 7.7 & 3.7 & 7.7 & 31.2\\
  & 0.9  & 0.002 & 1.7 & 7.4 & 0.079 & 7.8 & 3 & 7.8 & 34.5\\
  & 0.95 & 0.003 & 1.5 & 9.6 & 0.062 & 6.8 & 3 & 6.8 & 30.39\\
  \hline
\end{tabular}
\end{table}

\begin{table}
\centering 
\caption{\label{table9} The same as Table 2 except for $K^{*}\psi^{\prime}$.}
\vspace{0.2cm}
\tabcolsep=5pt
\begin{tabular}{llllllllll}
  \hline
  Reaction & $T/T_{\rm c}$ & $a_1$ & $b_1$ & $c_1$ & $a_2$ & $b_2$ & $c_2$ 
&$d_0$ & $\sqrt{s_{\rm z}}$\\
  \hline
$K^{*}+{\psi}^{\prime} \to D^{+}+X$
  & 0 & 0.01 & 2 & 5.1 & 0.31 & 8.4 & 3.5 & 8.4 & 34.78\\
  & 0.65 & 0.2 & 5.2 & 8.9 & 0.7 & 9.6 & 5.2 & 8.87 & 32.91\\
  & 0.75 & 0.1 & 4 & 9.5 & 0.6 & 7.5 & 8.1 & 7.36 & 22.91\\
  & 0.85 & 0.02 & 3 & 11 & 0.28 & 6.7 & 6.8 & 6.69 & 22.01\\
  & 0.9  & 0.002 & 2.2 & 11.2 & 0.164 & 6.9 & 4.8 & 6.9 & 25.4\\
  & 0.95 & 0.01 & 6 & 3 & 0.09 & 8.6 & 2.9 & 8.28 & 37.66\\
$K^{*}+{\psi}^{\prime} \to D^{0}+X$
  & 0 & 0.1 & 2.9 & 4.2 & 0.78 & 7.5 & 6 & 7.42 & 25.53\\
  & 0.65 & 0.08 & 3.5 & 24 & 1.91 & 7.5 & 6.3 & 7.5 & 25\\
  & 0.75 & 0.2 & 3.8 & 9.8 & 1.6 & 7.1 & 8.7 & 7.01 & 21.46\\
  & 0.85 & 0.1 & 3.3 & 9.5 & 0.8 & 6.9 & 8.5 & 6.86 & 21\\
  & 0.9  & 0.01 & 3 & 10 & 0.44 & 7.5 & 4.4 & 7.5 & 28.21\\
  & 0.95 & 0.001 & 1.2 & 7.7 & 0.229 & 7.3 & 3.2 & 7.3 & 31.38\\
$K^{*}+{\psi}^{\prime} \to D^{+}_{s}+X$
  & 0 & 0.005 & 1.7 & 6 & 0.151 & 7.1 & 3.7 & 7.1 & 29.48\\
  & 0.65 & 0.08 & 4 & 14 & 0.36 & 7 & 6.9 & 6.79 & 22.97\\
  & 0.75 & 0.06 & 3.5 & 15.7 & 0.31 & 6.5 & 8.8 & 6.45 & 20.04\\
  & 0.85 & 0.031 & 3.4 & 15.8 & 0.145 & 6.4 & 8.9 & 6.36 & 19.63\\
  & 0.9  & 0.009 & 3.1 & 16.7 & 0.083 & 6.4 & 6.9 & 6.39 & 21.12\\
  & 0.95 & 0.005 & 3.1 & 7.5 & 0.042 & 6.7 & 5.1 & 6.62 & 24.17\\
$K^{*}+{\psi}^{\prime} \to D^{*+}+X$
  & 0 & 0.01 & 2 & 6.3 & 0.32 & 7.7 & 4.7 & 7.7 & 28.69\\
  & 0.65 & 0.3 & 4.9 & 11.5 & 0.8 & 8 & 12.8 & 7.58 & 21.3\\
  & 0.75 & 0.1 & 4.2 & 16.3 & 0.7 & 7.5 & 7.5 & 7.43 & 23.58\\
  & 0.85 & 0.08 & 4 & 9 & 0.37 & 7.9 & 6.7 & 7.72 & 25.33\\
  & 0.9  & 0.04 & 4 & 6 & 0.17 & 7.5 & 4.8 & 6.96 & 27.12\\
  & 0.95 & 0.003 & 1.9 & 7.2 & 0.109 & 7.3 & 3 & 7.3 & 32.34\\
  \hline
\end{tabular}
\end{table}

\begin{table}
\centering 
\caption{\label{table10} The same as Table 2 except for $K^{*}\chi_c$.}
\vspace{0.2cm}
\tabcolsep=5pt
\begin{tabular}{llllllllll}
  \hline
  Reaction & $T/T_{\rm c}$ & $a_1$ & $b_1$ & $c_1$ & $a_2$ & $b_2$ & $c_2$ 
&$d_0$ & $\sqrt{s_{\rm z}}$\\
  \hline
$K^{*}+{\chi}_{c} \to D^{+}+X$
  & 0 & 0.02 & 3 & 3.5 & 0.3 & 9.1 & 3.9 & 9.05 & 35.54\\
  & 0.65 & 0.004 & 2 & 6.6 & 0.472 & 7.5 & 5.1 & 7.5 & 26.98\\
  & 0.75 & 0.1 & 5.5 & 7.4 & 0.4 & 8.2 & 4.7 & 7.4 & 29.57\\
  & 0.85 & 0.009 & 2.6 & 12.4 & 0.277 & 6.7 & 6 & 6.7 & 22.97\\
  & 0.9  & 0.01 & 3 & 7.3 & 0.18 & 7.8 & 4.6 & 7.79 & 28.66\\
  & 0.95 & 0.01 & 3.1 & 3.6 & 0.11 & 8.4 & 3.8 & 8.3 & 32.86\\
$K^{*}+{\chi}_{c} \to D^{0}+X$
  & 0 & 0.1 & 4.1 & 3.2 & 0.88 & 10.7 & 3.9 & 10.49 & 40.95\\
  & 0.65 & 0.02 & 6.7 & 2.1 & 1.23 & 7.7 & 4.9 & 7.69 & 28.24\\
  & 0.75 & 0.1 & 3.8 & 6.8 & 1.3 & 7.6 & 6.3 & 7.52 & 25.17\\
  & 0.85 & 0.02 & 2.4 & 14.9 & 0.75 & 6.9 & 5.9 & 6.9 & 23.67\\
  & 0.9  & 0.04 & 4.7 & 7.7 & 0.43 & 8 & 4 & 7.7 & 30.94\\
  & 0.95 & 0.01 & 4 & 3 & 0.29 & 8.7 & 3.1 & 8.59 & 37.16\\
$K^{*}+{\chi}_{c} \to D^{+}_{s}+X$
  & 0 & 0.02 & 3.3 & 3.8 & 0.14 & 8.1 & 4.2 & 7.9 & 31.22\\
  & 0.65 & 0.01 & 5 & 3 & 0.24 & 6.3 & 5.7 & 6.27 & 22.62\\
  & 0.75 & 0.04 & 4 & 10 & 0.25 & 6.9 & 6.1 & 6.64 & 23.57\\
  & 0.85 & 0.018 & 3.1 & 17.5 & 0.152 & 6.3 & 7.7 & 6.29 & 20.25\\
  & 0.9  & 0.02 & 3.4 & 11.2 & 0.08 & 6.4 & 9.6 & 6.28 & 19.09\\
  & 0.95 & 0.006 & 3.1 & 9.4 & 0.048 & 6.6 & 5.3 & 6.55 & 23.54\\
$K^{*}+{\chi}_{c} \to D^{*+}+X$
  & 0 & 0.01 & 2.7 & 5.2 & 0.31 & 8.6 & 4.3 & 8.6 & 32.64\\
  & 0.65 & 0.1 & 4.8 & 13.7 & 0.95 & 12.9 & 4 & 12.9 & 47.98\\
  & 0.75 & 0.1 & 4.7 & 10 & 0.5 & 8 & 5.9 & 7.56 & 26.9\\
  & 0.85 & 0.04 & 3.5 & 9.8 & 0.337 & 7.1 & 6.3 & 7.04 & 23.71\\
  & 0.9  & 0.02 & 3 & 6 & 0.19 & 6.7 & 5.9 & 6.63 & 22.97\\
  & 0.95 & 0.01 & 2.8 & 4.1 & 0.12 & 7.8 & 3.2 & 7.73 & 33.26\\
  \hline
  \end{tabular}
\end{table}

\begin{table*}[htbp]
\tabcolsep=5pt
\caption{\label{table11} The same as Table 2 except for $\pi J/{\psi}$.}
\vspace{0.2cm}
\begin{tabular}{*{13}{c}}
  \hline
Reaction & $T/T_{c}$ & $a_{1}$ & $b_{1}$ & $c_{1}$ & $a_{2}$ & $b_{2}$ 
& $c_{2}$ & $d_{0}$ & $\sqrt{s_{z}}$\\
  \hline
$\pi+ J/{\psi} \to D^{+}+X$
  & 0     & 0.01  & 0.01  & 0.09  & 0.12  & 9.27  & 4.18  & 9.27  & 34.91\\
  & 0.65  & 0.01  & 0.01  & 0.09  & 0.12  & 9.17  & 4.16  & 9.17  & 34.56\\
  & 0.75 & 0.01 & 0.01 & 0.1 & 0.11 & 8.67 & 4.16 & 8.67 & 32.82\\
  & 0.85 & 0.01 & 0.01 & 0.09 & 0.102 & 8.23 & 4.12 & 8.23 & 31.36\\
  & 0.9 & 0.01 & 0.01 & 0.11 & 0.099 & 8.08 & 3.93 & 8.08 & 31.35\\
  & 0.95 & 0.01 & 3.2 & 4.9 & 0.1 & 8.3 & 4.6 & 8.24 & 29.84\\
$\pi+ J/{\psi} \to D^{0}+X$ 
  & 0     & 0.01  & 0.01  & 0.09  & 0.3   & 9.15  & 4.23  & 9.15  & 34.34\\
  & 0.65  & 0.01  & 0.01  & 0.09  & 0.31  & 9.13  & 4.2   & 9.13  & 34.29\\
  & 0.75 & 0.01 & 0.01 & 0.08 & 0.3 & 8.92 & 4.16 & 8.92 & 33.65 \\
  & 0.85 & 0.01 & 3.1 & 5.4 & 0.29 & 8.6 & 4.7 & 8.59 & 30.86 \\
  & 0.9 & 0.01 & 0.01 & 0.09 & 0.27 & 8.33 & 3.93 & 8.33 & 32.19 \\
  & 0.95 & 0.02 & 3 & 5 & 0.26 & 8.1 & 4.6 & 8.07 & 29.21 \\
$\pi+ J/{\psi} \to D^{+}_{s}+X$ 
  & 0     & 0.01  & 0.01  & 0.1  & 0.06  & 7.98  & 4.41  & 7.98  & 30.1\\
  & 0.65  & 0.01  & 0.01  & 0.1  & 0.06  & 7.87  & 4.24  & 7.87  & 30.11\\
  & 0.75 & 0.01 & 4.5 & 5.1 & 0.06 & 9.5 & 4.5 & 9.14 & 34.41 \\
  & 0.85 & 0.01 & 4.1 & 5.4 & 0.05 & 8 & 6.2 & 7.67 & 26.13 \\
  & 0.9 & 0.01 & 4.2 & 5.7 & 0.05 & 8.7 & 5.2 & 8.36 & 29.88 \\
  & 0.95 & 0.006 & 3.5 & 6.3 & 0.044 & 7.6 & 5 & 7.46 & 26.88 \\
$\pi+ J/{\psi} \to D^{*+}+X$ 
  & 0     & 0.01  & 0.01  & 0.1   & 0.12  & 8.58  & 5.01  & 8.58  & 30.48\\
  & 0.65  & 0.01  & 0.01  & 0.09  & 0.12  & 8.3   & 4.85  & 8.3   & 29.89\\
  & 0.75 & 0.01 & 3.6 & 5.9 & 0.12 & 7.9 & 6.7 & 7.84 & 25.35 \\
  & 0.85 & 0.02 & 4.5 & 4.8 & 0.13 & 9.4 & 5 & 9.09 & 32.49 \\
  & 0.9 & 0.01 & 3.5 & 4.9 & 0.12 & 8.3 & 4.4 & 8.21 & 30.6 \\
  & 0.95 & 0.01 & 3 & 5.1 & 0.11 & 7.8 & 4 & 7.74 & 29.96 \\
  \hline
\end{tabular}
\end{table*}

\begin{table*}[htbp]
\tabcolsep=5pt
\caption{\label{table12} The same as Table 2 except for $\pi{\psi}^{\prime}$.}
\vspace{0.2cm}
\begin{tabular}{*{13}{c}}
  \hline
Reaction & $T/T_{c}$ & $a_{1}$ & $b_{1}$ & $c_{1}$ & $a_{2}$ & $b_{2}$ 
& $c_{2}$ & $d_{0}$ & $\sqrt{s_{z}}$   \\
  \hline
$\pi+{\psi}^{\prime} \to D^{+}+X$ 
  & 0     & 0.01  & 0.01  & 0.08  & 0.17  & 8.07  & 4     & 8.07  & 31.53\\
  & 0.65  & 0.01  & 0.01  & 0.03  & 0.25  & 6.97  & 7.64  & 6.97  & 21.93\\
  & 0.75 & 0.03 & 4.2 & 13 & 0.246 & 8.2 & 6.9 & 8.15 & 25.82\\
  & 0.85 & 0.017 & 3.8 & 11 & 0.172 & 7.7 & 6.5 & 7.66 & 24.86\\
  & 0.9 & 0.05 & 4.9 & 6.4 & 0.17 & 10.6 & 6 & 10.21 & 33.48\\
  & 0.95 & 0.01 & 3.7 & 5.1 & 0.12 & 8.5 & 4 & 8.38 & 32.3\\
$\pi+{\psi}^{\prime} \to D^{0}+X$ 
  & 0     & 0.01  & 0.01  & 0.06  & 0.44  & 8.04  & 4.06  & 8.04  & 31.23\\
  & 0.65  & 0.01  & 0.01  & 0.02  & 0.64  & 6.91  & 7.88  & 6.91  & 21.57\\
  & 0.75 & 0.05 & 4 & 14 & 0.614 & 7.7 & 7.2 & 7.67 & 24.15\\
  & 0.85 & 0.05 & 4 & 10 & 0.458 & 7.9 & 6.6 & 7.83 & 25.26\\
  & 0.9 & 0.1 & 4.5 & 6.7 & 0.36 & 8.4 & 7.6 & 7.98 & 25.12\\
  & 0.95 & 0.1 & 4.8 & 4.5 & 0.31 & 9.7 & 5.6 & 8.95 & 31.43\\
$\pi+{\psi}^{\prime} \to D^{+}_{s}+X$
  & 0     & 0.01  & 0.01  & 0.06  & 0.09  & 7.26  & 4.05  & 7.26  & 28.76\\
  & 0.65  & 0.03  & 4.1   & 18.7  & 0.14  & 7.5   & 7.3   & 7.44  & 23.69\\ 
  & 0.75 & 0.023 & 3.8 & 24 & 0.117 & 6.9 & 7.9 & 6.88 & 21.52\\
  & 0.85 & 0.021 & 3.8 & 19.5 & 0.091 & 7.6 & 7.1 & 7.59 & 23.98\\
  & 0.9 & 0.018 & 3.9 & 15.1 & 0.073 & 8.1 & 6.2 & 8.07 & 26.39\\
  & 0.95 & 0.02 & 4.2 & 8.3 & 0.07 & 10.5 & 4.9 & 10.44 & 35.98\\
$\pi+{\psi}^{\prime} \to D^{*+}+X$ 
  & 0     & 0.01  & 0.01  & 0.01  & 0.19  & 8.13  & 4.8   & 8.13  & 29.61\\
  & 0.65  & 0.09  & 5     & 16.7  & 0.82  & 14.2  & 5.7   & 14.2  & 44.98\\
  & 0.75 & 0.095 & 4.9 & 14.5 & 0.732 & 14.3 & 5.7 & 14.3 & 45.15\\
  & 0.85 & 0.031 & 3.9 & 13 & 0.202 & 7.7 & 6.7 & 7.65 & 24.64\\
  & 0.9 & 0.03 & 3.8 & 7.7 & 0.16 & 7.7 & 6.1 & 7.52 & 25.28\\
  & 0.95 & 0.03 & 4 & 4.9 & 0.15 & 9.8 & 3.8 & 9.52 & 37.49\\
  \hline
\end{tabular}
\end{table*}

\begin{table*}[htbp]
\tabcolsep=5pt
\caption{\label{table13} The same as Table 2 except for $\pi{\chi}_{c}$.}
\vspace{0.2cm}
\begin{tabular}{*{13}{c}}
  \hline
Reaction & $T/T_{c}$ & $a_{1}$ & $b_{1}$ & $c_{1}$ & $a_{2}$ & $b_{2}$ 
& $c_{2}$ & $d_{0}$ & $\sqrt{s_{z}}$   \\
  \hline
$\pi+{\chi}_{c} \to D^{+}+X$ 
  & 0     & 0.01  & 0.01  & 0.1   & 0.17  & 8.86  & 3.72  & 8.86  & 35.27\\
  & 0.65  & 0.01  & 0.01  & 0.06  & 0.21  & 7.83  & 5.29  & 7.83  & 27.41\\
  & 0.75 & 0.05 & 5 & 7.2 & 0.21 & 8.8 & 6.9 & 8.32 & 27.27\\
  & 0.85 & 0.02 & 4 & 10 & 0.18 & 8.2 & 6 & 8.14 & 26.97\\
  & 0.9 & 0.03 & 4.4 & 6.7 & 0.15 & 9 & 5.6 & 8.71 & 29.79\\
  & 0.95 & 0.01 & 3.8 & 5.8 & 0.13 & 8.8 & 3.8 & 8.71 & 34.07\\
$\pi+{\chi}_{c} \to D^{0}+X$
  & 0     & 0.01  & 0.01  & 0.07  & 0.45  & 8.98  & 3.75  & 8.98  & 35.56\\
  & 0.65  & 0.01  & 0.01  & 0.05  & 0.54  & 7.85  & 5.33  & 7.85  & 27.39\\ 
  & 0.75 & 0.1 & 5 & 9 & 0.68 & 10.3 & 5 & 10.14 & 35.38\\
  & 0.85 & 0.12 & 4.7 & 8 & 0.495 & 9.2 & 6.9 & 8.9 & 28.26\\
  & 0.9 & 0.1 & 4.7 & 6.5 & 0.41 & 9.6 & 5.8 & 9.21 & 31.07\\
  & 0.95 & 0.1 & 5 & 4.5 & 0.36 & 11 & 4.4 & 10.26 & 38.89\\
$\pi+{\chi}_{c} \to D^{+}_{s}+X$ 
  & 0     & 0.01  & 0.01  & 0.09  & 0.09  & 7.74  & 3.93  & 7.74  & 30.74\\
  & 0.65  & 0.01  & 0.01  & 0.06  & 0.11  & 6.6   & 5.85  & 6.6   & 23.01\\
  & 0.75 & 0.01 & 3.6 & 19 & 0.109 & 6.94 & 6.42 & 6.93 & 23.14\\
  & 0.85 & 0.02 & 3.8 & 15 & 0.085 & 7.1 & 8.2 & 7.02 & 21.65\\
  & 0.9 & 0.02 & 3.9 & 12.7 & 0.07 & 7.6 & 7.7 & 7.49 & 23.23\\
  & 0.95 & 0.02 & 3.9 & 9.3 & 0.06 & 8 & 7.6 & 7.81 & 24.15\\
$\pi+{\chi}_{c} \to D^{*+}+X$ 
  & 0     & 0.01  & 0.01  & 0.08  & 0.17  & 8.09  & 4.64  & 8.09  & 29.88\\
  & 0.65  & 0.01  & 0.01  & 0.07  & 0.23  & 7.96  & 5.92  & 7.96  & 26.77\\  
  & 0.75 & 0.08 & 5.3 & 9.8 & 0.377 & 12.5 & 5.2 & 12.45 & 41.48\\
  & 0.85 & 0.027 & 3.8 & 11.6 & 0.199 & 7.5 & 6.9 & 7.44 & 23.87\\
  & 0.9 & 0.03 & 3.9 & 7.9 & 0.17 & 8.2 & 5.3 & 8.04 & 28.11\\
  & 0.95 & 0.03 & 4 & 5.1 & 0.15 & 9.4 & 3.8 & 9.08 & 36.1\\
  \hline
\end{tabular}
\end{table*}

\begin{table}
\centering \caption{\label{table14} Values selected from references are
peak cross sections, and the four values in the last row are the cross sections
obtained in the present work at $\sqrt{s}=11$ GeV and $T=0$ GeV.  The cross
sections are in units of mb.}
\vspace{0.2cm}
\begin{tabular}{|l|l|l|l|l|}
  \hline
  Reference & $\pi J/\psi$ & $\rho J/\psi$ & $K J/\psi$ & $K^* J/\psi$ \\
  \hline
  \cite{Haglin}  & 2.5   & 0.9   &       &  \\
  \hline
  \cite{LK}      & 3.7   & 2.9   &       &  \\
  \hline
  \cite{OSL}     & 6.2   & 3     &       &  \\
  \hline 
  \cite{AN}      &       &       & 1.56  &  \\
  \hline
  \cite{BG}      & 0.83  &       &       &  \\
  \hline
  \cite{CMR}     & 3.8   &       &       &  \\
  \hline
  \cite{AKM}     & 7     & 2.7   &       &  \\
  \hline
  \cite{ACM}     & 0.5   & 1.7   & 0.36  & 1.2 \\
  \hline
  \cite{MBQ}     & 7     &       &       &  \\
  \hline
  \cite{WSB}     & 0.9   & 5.7   & 0.8   &  \\
  \hline
  \cite{BSWX}    & 1.41  & 10.4  &       &  \\
  \hline
  \cite{ZX}      & 0.37  & 2.12  &       &  \\
  \hline
  \cite{JSX}     &       &       & 0.6   &  \\
  \hline
  \cite{LJX}     &       &       &       & 1.05 \\
  \hline
                 & 0.82  & 1.28  & 0.76  & 1.11 \\
  \hline
  \end{tabular}
\end{table}


\begin{thebibliography}{00}
\bibitem{KS}D. Kharzeev and H. Satz, Phys. Lett. B 334, 155 (1994).
\bibitem{Peskin}M. E. Peskin, Nucl. Phys. B 156, 365 (1979); G. Bhanot 
and M. E. Peskin, Nucl. Phys. B 156, 391 (1979).
\bibitem{AGGA}F. Arleo, P. B. Gossiaux, T. Gousset, and J. Aichelin, Phys. Rev.
D 65, 014005 (2001).
\bibitem{MM}S. G. Matinyan and B. M\"{u}ller, Phys. Rev. C 58, 2994
(1998). 
\bibitem{Haglin}K. L. Haglin, Phys. Rev. C 61, 031902(R) (2000); 
K. L. Haglin and C. Gale, Phys. Rev. C 63, 065201 (2001).
\bibitem{LK}Z. Lin and C. M. Ko, Phys. Rev. C 62, 034903 (2000); J. Phys. G 27,
617 (2001).
\bibitem{OSL}Y. S. Oh, T. S. Song, and S. H. Lee, Phys. Rev. C 63, 034901
(2001).
\bibitem{NNR}F. S. Navarra, M. Nielsen, and M. R. Robilotta, Phys. Rev. C 
64, 021901(R) (2001).
\bibitem{MPPR} L. Maiani, F. Piccinini, A. D. Polosa, and V. Riquer, Nucl.
Phys. A 741, 273 (2004).
\bibitem{AN}R. S. Azevedo and M. Nielsen, Phys. Rev. C 69, 035201 (2004).
\bibitem{BG}A. Bourque and C. Gale, Phys. Rev. C 80, 015204 (2009).
\bibitem{CMR}M. Cleven, V. K. Magas, and A. Ramos, Phys. Rev. C 96, 045201 
(2017).
\bibitem{AKM}L. M. Abreu, K. P. Khemchandani, A. M. Torres, F. S. Navarra, and
M. Nielsen, Phys. Rev. C 97, 044902 (2018).
\bibitem{ACM}L. M. Abreu, E. Cavalcanti, and A. P. C. Malbouisson, Nucl. Phys.
A 978, 107 (2018).
\bibitem{BS}T. Barnes and E. S. Swanson, Phys. Rev. D 46, 131 (1992); E. S. 
Swanson, Ann. Phys. (N.Y.) 220, 73 (1992).
\bibitem{MBQ}K. Martins, D. Blaschke, and E. Quack, Phys. Rev. C 51, 
2723 (1995).
\bibitem{WSB}C.-Y. Wong, E. S. Swanson, and T. Barnes, Phys. Rev. C 62, 045201
(2000); Phys. Rev. C 65, 014903 (2001).
\bibitem{BSWX}T. Barnes, E. S. Swanson, C.-Y. Wong, and X.-M. Xu, Phys. Rev. C
68, 014903 (2003).
\bibitem{ZX}J. Zhou and X.-M. Xu, Phys. Rev. C 85, 064904 (2012).
\bibitem{JSX}S.-T. Ji, Z.-Y. Shen, and X.-M. Xu, J. Phys. G 42, 095110 (2015).
\bibitem{LJX}F.-R. Liu, S.-T. Ji, and X.-M. Xu, J. Korean Phys.
Soc. 69, 472 (2016).
\bibitem{DKLNN}F. O. Dur\~{a}es, H. Kim, S. H. Lee, F. S. Navarra, and 
M. Nielsen, Phys. Rev. C 68, 035208 (2003).
\bibitem{Wong}C.-Y. Wong, Phys. Rev. C 65, 034902 (2002).
\bibitem{Xu2002}X.-M. Xu, Nucl. Phys. A 697, 825 (2002).
\bibitem{AV}L. M. Abreu and H. P. L. Vieira, Eur. Phys. J. A 57, 163 (2021).
\bibitem{CMS}A. M. Sirunyan {\it et al.}, Eur. Phys. J. C 78, 509 (2018).
\bibitem{ATLAS}M. Aaboud {\it et al.}, Eur. Phys. J. C 78, 762 (2018).
\bibitem{JXH}S.-T. Ji, X.-M. Xu, and H. J. Weber, Nucl. Phys. A 966, 224 
(2017).
\bibitem{FF}R. D. Field and R. P. Feynman, Nucl. Phys. B 136, 1 (1978); R. D. 
Field, {\it Applications of Perturbative QCD} (Addison-Wesley, Redwood City, 
CA, 1989).
\bibitem{BT}W. Buchm\"{u}ller and S.-H. H. Tye, Phys. Rev. D 24, 132 (1981).
\bibitem{GI}S. Godfrey and N. Isgur, Phys. Rev. D 32, 189 (1985).
\bibitem{BD}J. D. Bjorken and S. D. Drell, {\it Relativistic Quantum Mechanics}
(McGraw-Hill, New York, 1964).
\bibitem{Schwabl}F. Schwabl, {\it Quantum Mechanics} (Springer-Verlag, Berlin,
2007).
\bibitem{KK}B. A. Kniehl and G. Kramer, Phys. Rev. D 74, 037502 (2006).
\bibitem{KKKS}T. Kneesch, B. A. Kniehl, G. Kramer,
and I. Schienbein, Nucl. Phys. B 799, 34 (2008).
\bibitem{Chen}J.-Q. Chen, {\it A New Apporach to Group Representations}
(Shanghai Scientific and Technical Publishers, Shanghai, 1984).
\end{thebibliography}
\end{document}